\documentclass[12pt,a4paper]{article}

\usepackage{amsmath}
\usepackage{amssymb}
\usepackage{graphicx}
\usepackage{dsfont}
\usepackage{overpic}
\usepackage{cite}
\usepackage{hhline}
\usepackage{dcolumn}

\let\oldappendix=\appendix
\let\oldsection=\section
\renewcommand{\appendix}{\oldappendix%
\def\theequation{\Alph{section}.\arabic{equation}}%
\renewcommand{\section}{\setcounter{equation}{0}\oldsection}}

\newcommand{\beq}{\begin{equation}}
\newcommand{\eeq}{\end{equation}}
\newcommand{\beqa}{\begin{eqnarray}}
\newcommand{\eeqa}{\end{eqnarray}}

\newcommand{\tr}{\mbox{tr}}
\newcommand{\sfrac}[2]{{\textstyle\frac{#1}{#2}}}
\newcommand{\pn}{\tilde{\pi}^0}
\newcommand{\deltaph}{\lambda}
\def\bra#1{\left\langle #1\right|}
\def\ket#1{\left| #1\right\rangle}
\def\trf#1{\left\langle #1 \right\rangle}

\newcommand{\newop}[2]{\def#1{\mathop{\mathrm{#2}}\nolimits}}
\newop{\artanh}{artanh}
\newop{\det}{det}
\newop{\tr}{tr}
\newop{\diag}{diag}
\newop{\Re}{Re}
\newop{\Im}{Im}

\newcommand{\mmass}[2][2]{{m_{#2}^{#1}}}
\newcommand{\masseps}[1][2]{\mmass{\epsilon}}
\newcommand{\masstop}[1][2]{\mmass{0}}

\newcommand{\tsum}{\mathop{\textstyle\sum}}

\newcommand{\decay}{f}
\newcommand{\coeffv}[2]{v_{#1}^{(#2)}}
\newcommand{\cbeta}[2]{\beta_{#1}^{(#2)}}
\newcommand{\cvtwid}[2]{\tilde{v}_{#1}^{(#2)}}

\newcommand{\eV}{\,\mathrm{eV}}
\newcommand{\keV}{\,\mathrm{keV}}
\newcommand{\MeV}{\,\mathrm{MeV}}
\newcommand{\GeV}{\,\mathrm{GeV}}
\newcommand{\Lagr}{\mathcal{L}}

\newcommand{\indup}[1]{_{\scriptscriptstyle\mathrm{#1}}}

\begin{document}

\hfill 

\hfill 

\bigskip\bigskip

\begin{center}

{{\Large\bf  Hadronic $\mbox{\boldmath$\eta$}$ and $\mbox{\boldmath$\eta'$}$ decays}}

\end{center}

\vspace{.4in}

\begin{center}
{\large B.~Borasoy\footnote{email: borasoy@itkp.uni-bonn.de} and
        R.~Ni{\ss}ler\footnote{email: rnissler@itkp.uni-bonn.de}}

\bigskip

\bigskip

Helmholtz-Institut f\"ur Strahlen- und Kernphysik (Theorie) \\
Universit\"at Bonn \\ 
Nu{\ss}allee 14-16, D-53115 Bonn, Germany

\vspace{.2in}

\end{center}

\vspace{.7in}

\thispagestyle{empty} 

\begin{abstract}

The hadronic decays $\eta, \eta' \to 3 \pi$ and $\eta' \to \eta \pi \pi$
are investigated within the framework of U(3) chiral effective field theory
in combination with a relativistic coupled-channels approach.
Final state interactions are included by deriving $s$- and $p$-wave interaction
kernels for meson-meson scattering from the chiral effective Lagrangian and
iterating them in a Bethe-Salpeter equation.
Very good overall agreement with currently available data on decay widths and 
spectral shapes is achieved.

\end{abstract}\bigskip

\begin{center}
\begin{tabular}{ll}
\textbf{PACS:}& 12.39.Fe, 13.25.Jx \\[6pt]
\textbf{Keywords:}& Chiral Lagrangians, coupled channels, $\eta$ and $\eta'$.
\end{tabular}
\end{center}


\vfill

\section{Introduction}\label{sec:Intro}

The hadronic decays of $\eta$ and $\eta'$ offer a possibility to study symmetries
and symmetry breaking patterns in strong interactions.
The isospin-violating decays $\eta, \eta' \to 3 \pi$, e.g., can only occur
due to an isospin-breaking quark mass difference $m_u-m_d$ or electromagnetic
effects. While for most processes isospin-violation of the strong interactions is
masked by electromagnetic effects, these corrections are expected to be small for the
three pion decays of $\eta$ and $\eta'$ (Sutherland's theorem) \cite{Sutherland:1966mi} 
which has been confirmed in an effective Lagrangian framework \cite{BKW}.
Neglecting electromagnetic corrections the decay amplitude is directly proportional to 
$m_u-m_d$.

Moreover, the $\eta$-$\eta'$ system offers a testing ground for chiral 
SU(3) symmetry in QCD and the role of both spontaneous and explicit chiral symmetry breaking,
the latter one induced by the light quark masses.
In the absence of $\eta$-$\eta'$ mixing, $\eta$ would be the pure member $\eta_8$ of the octet
of Goldstone bosons which arise due to spontaneous breakdown of chiral symmetry.

Reactions involving the $\eta'$
might also provide insight into gluonic effects through the axial U(1) anomaly of QCD.
The divergence of the singlet axial-vector current acquires an additional term
with the gluonic field strength tensor that remains
in the chiral limit of vanishing light quark masses.
This term prevents the pseudoscalar
singlet $\eta_0$ from being a Goldstone boson which is phenomenologically
manifested in its relatively large mass, $m_{\eta'} = 958$ MeV.

An appropriate theoretical framework to investigate low-energy hadronic physics is
provided by chiral perturbation theory (ChPT) \cite{GL}, the effective field theory 
of QCD. In ChPT Green's functions are expanded perturbatively in powers of Goldstone boson
masses and small three-momenta. 
However, final state interactions in $\eta \to 3 \pi$ 
have been shown to be substantial both in a complete one-loop calculation
in SU(3) ChPT \cite{GL2} and using extended Khuri-Treiman equations \cite{Kambor}.

In $\eta'$ decays final state interactions are expected to be even more important
due to larger phase space and the presence of nearby resonances.
It is claimed, e.g., that the exchange of the scalar resonance $a_0(980)$ dominates
the decays $\eta' \to \eta \pi \pi$ \cite{Fariborz} which has been
confirmed both in a full one-loop calculation utilizing infrared regularization \cite{BB2}
and in a chiral unitary approach \cite{BB1}.
In the latter work, resonances are generated dynamically by iterating 
the chiral effective potentials to infinite order in a BSE, whereas in \cite{BB2}
the effects of the $a_0(980)$ are hidden in a combination of coupling constants
of the effective Lagrangian.

In the present investigation we extend the approach of \cite{BB1} by including
$p$-wave interactions. This will also allow us
to obtain more realistic predictions for the decay $\eta' \to \pi^+ \pi^- \pi^0$,
where $p$-waves can---in principle---yield sizable contributions to the decay width and 
Dalitz slope parameters.
Furthermore, we study the implications of two very recent experiments by the 
KLOE \cite{KLOE} and the VES \cite{VES} Collaborations
which have determined the Dalitz plot distributions of $\eta \to 3 \pi$ and $\eta' \to \eta \pi^+ \pi^-$,
respectively, with high statistics. Since these new data have not been published yet, we 
first present the results which we obtain by relying purely on the numbers quoted by the Particle Data
Group (PDG) \cite{pdg}. As a second step, we include both new experiments separately in the fit and
discuss the resulting changes.

The improved analysis of $\eta, \eta'$ hadronic decays presented here is also timely in view
of the planned WASA facility at COSY \cite{WASA} and MAMI-C \cite{Nef}
which will provide even higher statistics for these decays.
More precise data will help to constrain the parameters of the Lagrangian
and pose tighter constraints on the framework employed here.
We will illustrate that meson-meson scattering phase shifts
along with available data on $\eta, \eta'$ hadronic decays
provide a set of tight constraints which must be met by theoretical approaches.

This work is organized as follows. In the next section details of the effective Lagrangian
in the U(3) framework are given. Section~\ref{sec:fsi} illustrates our way of incorporating
final state interactions and includes a discussion of constraints set by unitarity.
In Sec.~\ref{sec:res} we present our results based on data from \cite{pdg}
and the changes which arise if the new, but preliminary experimental results by KLOE \cite{KLOE}
and VES \cite{VES} are included.
A critical examination of the $\eta \to 3 \pi$ data of KLOE 
based on purely phenomenological arguments is presented in Sec.~\ref{sec:DalPhe}.
We summarize our findings in Sec.~\ref{sec:con}.

\section{Effective Lagrangian}

In this section we present the effective Lagrangian within the framework of U(3) chiral perturbation theory
and summarize the resulting $\pn$-$\eta_8$-$\eta_0$ mixing \cite{BB1}. 
Up to second order in the derivative expansion
the Lagrangian for the nonet of pseudoscalar mesons 
($\pi^{+/-}, \pn, K^{+/-}, K^0, \bar{K}^0, \eta_8, \eta_0$) 
reads (note that we do not make use of large-$N_c$ counting rules) \cite{KL1, KL2, BW}
\beq \label{eq:Lagr02}
\mathcal{L}^{(0+2)} = - V_0
+V_1 \langle \partial_{\mu} U^{\dagger} \partial^{\mu}U \rangle
+V_2 \langle U^\dagger \chi+\chi^\dagger U\rangle
+i V_3 \langle U^\dagger \chi-\chi^\dagger U\rangle
+V_4 \langle U^\dagger\partial^{\mu} U \rangle \langle U^\dagger\partial_{\mu} U \rangle,
\eeq
where $U$ is a unitary $3 \times 3$ matrix which collects the pseudoscalar fields. Its dependence on $\pn$, 
$\eta_8$ and $\eta_0$ is given by
\beq
U = \exp{\left( \diag(1,-1,0) \frac{i \pn}{f} + \diag(1,1,-2) \frac{i \eta_8}{\sqrt{3}f}
                + \frac{i \sqrt{2}\eta_0}{\sqrt{3}f} + \cdots \right)}\ .
\eeq
The expression $\langle \dots \rangle$ denotes the trace in flavor space, $f$ is the pseudoscalar decay constant
in the chiral limit and the quark mass matrix $\mathcal{M} = \diag(m_u, m_d, m_s)$ enters in the combination
$\chi = 2 B \mathcal{M}$ with $B = -\bra{0} \bar{q}q \ket{0}/f^2$ being the order parameter of spontaneous chiral 
symmetry breaking.
The coefficients $V_i$ are functions of the singlet field $\eta_0$ and can be expanded in terms of this variable
\begin{eqnarray} \label{eq:vexpand}
V_i\Big[\frac{\eta_0}{\decay}\Big] &=& \coeffv{i}{0} + \coeffv{i}{2}
\frac{\eta_0^2}{\decay^2} +
\coeffv{i}{4} \frac{\eta_0^4}{\decay^4} + \ldots
\qquad \mbox{for} \quad i= 0,1,2,4 \nonumber \\
V_3\Big[\frac{\eta_0}{\decay}\Big] &=& \coeffv{3}{1} \frac{\eta_0}{\decay} +
\coeffv{3}{3} \frac{\eta_0^3}{\decay^3}
+ \ldots \quad
\end{eqnarray}
with expansion coefficients $v_{i}^{(j)}$ not fixed by chiral symmetry. Parity conservation implies that the $V_i$ 
are all even functions of $\eta_0$ except $V_3$, which is odd, and 
$V_1(0) = V_2(0) = V_1(0) - 3 V_4(0) = \frac{1}{4} f^2$ gives the correct normalization for the quadratic terms of 
the mesons. Thus, at a given order of the derivative expansion one obtains an infinite string of increasing powers of 
$\eta_0$ preceding each term in the Lagrangian.

In order to describe the isospin-violating decays $\eta, \eta' \to 3 \pi$, we need to work with different up- and
down-quark masses and their difference $m_u - m_d$ will be parametrized by means of the 
renormalization scale invariant quantity
\beq \label{eq:meps}
m_{\epsilon}^{2} = B (m_d - m_u)
\eeq
which can be expressed in terms of physical meson masses following Dashen's theorem \cite{Dashen}
\beq \label{eq:Dashen}
m_{\epsilon}^{2} = m_{K^0}^{2} - m_{K^\pm}^{2} + m_{\pi^\pm}^{2} - m_{\pi^0}^{2}
\eeq
up to corrections of $\mathcal{O}(e^2 p^2, (m_d - m_u) p^2)$.

While in the isospin limit of equal up- 
and down-quark masses $\pn$ does not undergo mixing with the $\eta_8$-$\eta_0$ system, 
a non-vanishing quark mass difference $m_u - m_d$ induces $\pn$-$\eta_8$-$\eta_0$ mixing.
The mixing of $\pn$, $\eta_8$ and $\eta_0$ is determined by diagonalizing both the kinetic and mass 
terms in the Lagrangian. Since we count the mass of the $\eta'$ as a quantity of zeroth chiral order, 
the $\mathcal{O}(p^4)$ Lagrangian 
\beq
\Lagr^{(4)} = \sum_{i} \beta_i (\eta_0) \mathcal{O}_i 
\eeq
contributes to $\eta_8$-$\eta_0$ mixing already at leading chiral
order \cite{BB3}. The operators $\mathcal{O}_i$ can be found
in \cite{H-S, BB3},
but for completeness we will display in Appendix~\ref{sec:AppLagr} those relevant for the present calculation. 
The functions $\beta_i$ can be expanded in $\eta_0$ in the same manner as the $V_i$ in 
Eq.~(\ref{eq:Lagr02}) with expansion coefficients $\cbeta{i}{j}$.
At leading order in isospin-breaking and at second chiral order the mass eigenstates
$\pi^0$, $\eta$, $\eta'$ are related to the original fields $\pn$, $\eta_8$, $\eta_0$ by \cite{BB1}
\begin{eqnarray}\label{eq:mix4}
\pn &=&  (1+R_{\pn\pi^0}) \pi^0 + R_{\pn\eta}  \eta + R_{\pn\eta'}  \eta' \nonumber \\
\eta_8 &=&  R_{8\pi^0} \pi^0 + (1+R_{8\eta})  \eta + R_{8\eta'}  \eta' \nonumber \\
\eta_0 &=&  R_{0\pi^0} \pi^0 + R_{0\eta}  \eta + (1+R_{0\eta'})  \eta'
\end{eqnarray}
with the mixing parameters given by
\begin{equation} \label{eq:mixpar}
\begin{array}{rlcrl}
R_{8\pi^0}^{(0)}&=\displaystyle\frac{\masseps}{\sqrt{3}(\mmass{\eta}-\mmass{\pi})},&\qquad&
  R_{\pn\eta}^{(0)}&=-R_{8\pi^0}^{(0)},\\[15pt]
R_{8\pi^0}^{(2)}&=R_{8\pi^0}^{(0)} \Bigl(R_{\pn\pi^0}^{(2)}+\sfrac{2}{3}\Delta\indup{GMO}\Bigr),&&
  R_{\pn\eta}^{(2)}&=-R_{8\pi^0}^{(0)} \Bigl(R_{8\eta}^{(2)}+\sfrac{2}{3}\Delta\indup{GMO}\Bigr),\\[10pt]
R_{0 \eta}^{(2)}&=\displaystyle\frac{4\cvtwid{2}{1}(\mmass{\eta}-\mmass{\pi})}{\sqrt{2}f^2 \masstop},&&
  R_{8 \eta'}^{(2)}&=\displaystyle-R_{0 \eta}^{(2)}+\frac{8\cbeta{5,18}{0}(\mmass{\eta}-\mmass{\pi})}{\sqrt{2}f^2},\\[15pt]
R_{0\pi^0}^{(2)}&=3R_{8\pi^0}^{(0)}R_{0 \eta}^{(2)},&&
  R_{\pn\eta'}^{(2)}&=2R_{8\pi^0}^{(0)}R_{8 \eta'}^{(2)},  \\[5pt]
R_{\pn\pi^0}^{(2)}&=\displaystyle-\frac{4 \cbeta{5}{0} \mmass{\pi}+6 \cbeta{4}{0}(\mmass{\eta}+\mmass{\pi})}{f^2},&&
  R_{0 \eta'}^{(2)}&=\displaystyle-\frac{2\cbeta{4,5,17,18}{0}(\mmass{\eta}+\mmass{\pi})}{f^2},\\[12pt]
R_{8\eta}^{(2)}&=\displaystyle-\frac{4 \cbeta{5}{0} \mmass{\eta}+6 \cbeta{4}{0}(\mmass{\eta}+\mmass{\pi})}{f^2},
\end{array}
\end{equation}
where the superscript on $R$ denotes the chiral order and we have employed the
abbreviations $\cvtwid{2}{1} = \frac{1}{4}f^2-\frac{1}{2}\sqrt{6}\coeffv{3}{1}$,
$\cbeta{5,18}{0}= \cbeta{5}{0} + 3 \cbeta{18}{0}/2$
and $\cbeta{4,5,17,18}{0}= 3 \cbeta{4}{0} + \cbeta{5}{0} - 9 \cbeta{17}{0} + 3 \cbeta{18}{0}$.
Note that for $\cbeta{5,18}{0} \neq 0$ the mixing parameters given in Eq.~(\ref{eq:mixpar}), which 
have been derived by diagonalizing the Lagrangian, depart from the usually employed orthogonal mixing 
scheme for $\eta$-$\eta'$ mixing, where $R_{8 \eta'} = -R_{0 \eta}$.
Isospin-breaking is known to be small, therefore terms of order $(m_u - m_d)^2$ have been neglected.
The quantity $m^2_{0}=2v_0^{(2)}/f^2$ is the mass of the $\eta'$ meson in the chiral limit,
$m_\pi^2 = 2 B \hat{m}$ and $m_\eta^2 = 4 B (m_s + 2 \hat{m})/3$ 
with $\hat{m} = (m_u + m_d)/2$ denote
the pseudoscalar meson masses at leading order, while
the deviation from the Gell-Mann--Okubo mass relation for the pseudoscalar mesons is given by
\begin{equation}
\Delta\indup{GMO}=\frac{4\mmass{K}-\mmass{\pi}-3\mmass{\eta}}{\mmass{\eta}-\mmass{\pi}}=
\frac{6(\mmass{\eta}-\mmass{\pi})}{f^2}
\bigg[\cbeta{5}{0}-6\cbeta{8}{0}-12\cbeta{7}{0}+\frac{4(\cvtwid{2}{1})^2}{f^2 \masstop}\bigg] \ ,
\end{equation}
where $m_K^2= B(\hat{m} + m_s)$.

\section{Final state interactions}  \label{sec:fsi}

One-loop corrections have been shown to be important in the decay $\eta \to 3\pi$ \cite{GL2}, but even when 
they are taken into account
the resulting decay width remains below the measured value \cite{pdg}. 
In $\eta' \to 3 \pi$ one expects even larger contributions from final state interactions
\cite{BB1}, whereas for $\eta' \to \eta \pi \pi$
reasonable agreement with experiment can also be achieved in a perturbative approach employing infrared 
regularization \cite{BB2}. In the present combined analysis of these three dominant hadronic decay 
modes of $\eta$ and $\eta'$ we include final state interactions in a non-perturbative fashion as 
introduced in \cite{BB1}, but extending the work of \cite{BB1} by taking $p$-waves into account and by 
improving the fit procedure for the unknown couplings in the chiral Lagrangian via Monte Carlo techniques.

\begin{figure}
\centering\includegraphics[width=0.28\textwidth]{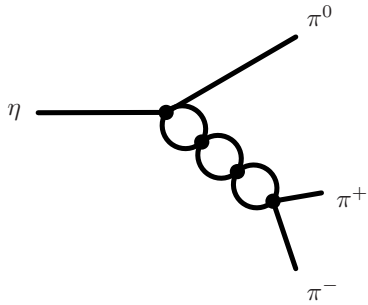}
\caption{Shown is a possible contribution to final state interactions in the decay $\eta \to \pi^+ \pi^- \pi^0$.}
\label{fig:Method}
\end{figure}

The underlying idea of our approach is that the initial particle, i.e.\ the $\eta$ or $\eta'$, decays into three 
mesons and that two out of these rescatter (elastically or inelastically) an arbitrary number of times, see 
Fig.~\ref{fig:Method} for illustration. All occurring vertices are derived from the effective Lagrangian and are 
thus constrained by chiral symmetry. 
Interactions of the third meson with the pair of rescattering mesons are neglected which turns out to be a 
good approximation, particularly for the decays $\eta \to 3\pi$ and $\eta' \to \eta \pi \pi$.
In the decays under consideration the two-particle states either carry one elementary charge or no net charge. 
Charge conservation prevents transitions between these two sets, while the different channels of one set are 
generally coupled. There are nine uncharged combinations of mesons
\begin{equation}
\pi^0\pi^0,\ \pi^+\pi^-,\ \eta\pi^0,\ \eta\eta,\
K^0\bar K^0,\ K^+K^-,\ \eta'\pi^0,\ \eta'\eta,\ \eta'\eta' \, ,
\end{equation}
and a set of four charged channels
\begin{equation}
\pi^0\pi^+,\ \eta\pi^+,\ K^+\bar K^0,\ \eta'\pi^+ \, .
\end{equation}
For the $p$-waves there arise some simplifications
in the uncharged channels. Due to Bose symmetry contributions from identical particles vanish and the remaining 
two-particle states can be classified according to their behavior under charge conjugation. While for $J = 1$ 
$\pi^+ \pi^-$ and $K \bar{K}$ must be $C$-odd combinations, the other pairs are $C$-even,
so that transitions between 
the two classes of states are forbidden.

For each partial-wave $l$ unitarity imposes a restriction on the (inverse) $T$-matrix of meson-meson scattering 
above threshold
\begin{equation} \label{unit}
\mbox{Im} T^{-1}_l = - \frac{|\mbox{\boldmath$q$}_{cm}|}{8 \pi E_{cm}}
\end{equation}
with $E_{cm}$ and $\mbox{\boldmath$q$}_{cm}$ being the energy and the three-momentum
of the particles in the center-of-mass (c.\,m.)\ frame of the channel under consideration, respectively.
Hence, the imaginary part of $T^{-1}_l$ is equal to the imaginary piece
of the fundamental scalar loop integral $\tilde{G}_{m \bar{m}}$ above threshold
\begin{equation}
\tilde{G}_{m \bar{m}}(p^2) =
\int\frac{\,d^d k}{(2\pi)^d}\,\frac{i}{(k^2-m^2+i \epsilon)( (k-p)^2-\bar m^2+i \epsilon)} \ ,
\end{equation}
with $m$, $\bar{m}$ denoting the masses of the two particles.
In dimensional regularization the finite part $G_{m \bar{m}}$ of $\tilde{G}_{m \bar{m}}$ is given by
{\arraycolsep0pt\begin{eqnarray}  \label{eq:g}
G_{m \bar{m}}(p^2) =&& \mathrel{} a_{m \bar{m}}(\mu) +
\frac{1}{16\pi^2}\bigg[-1+
\ln\frac{m \bar{m}}{\mu^2}
  +\frac{m^2-\bar{m}^2}{p^2}\ln\frac{m}{\bar{m}}
\nonumber\\&&\qquad\qquad\qquad\qquad\quad\mathord{}
  -\frac{2\sqrt{\deltaph_{m\bar{m}}(p^2)}}{p^2}\artanh\frac{
  \sqrt{\deltaph_{m\bar{m}}(p^2)}}{(m+\bar{m})^2-p^2}\bigg] \ ,
\nonumber\\
\deltaph_{m\bar{m}}(p^2)=&&\mathrel{}\big((m-\bar{m})^2-p^2\big)\big((m+\bar{m})^2-p^2\big) \ ,
\end{eqnarray}}%
where $a_{m \bar{m}}$ is a subtraction constant which varies with the scale $\mu$ introduced in dimensional 
regularization in such a way that $G_{m \bar{m}}$ is scale-independent \cite{OM}.
While unitarity constrains the imaginary part of the inverse $T$-matrix, the real part can be linked to ChPT. 
From the effective Lagrangian up to fourth chiral
order we derive the c.\,m.\ scattering amplitude $A_{fi}(\theta)$, where $\theta$ is the c.\,m.\ scattering angle
and the indices $i$, $f$ denote the in- and out-going meson pairs, respectively.
The decomposition into partial waves reads
\beq
A(\theta) = \sum_{l=0}^{2} (2l+1) A_l \,P_l(\cos \theta) \ .
\eeq
As usual $P_l$ denotes the $l^{\textrm{th}}$ Legendre polynomial.
Note that partial waves with $l>2$ do not occur, since we consider the effective Lagrangian only 
up to fourth chiral order.
Given a set of $n$ coupled channels the partial-wave amplitudes $A_l$ (just as the full amplitude $A$) are 
represented by symmetric $n \times n$ matrices which are functions of the c.\,m.\ energy $E_{cm}$.
The symmetry factor for two identical particles in a loop is absorbed into the $A_{fi}$ by equipping 
the matrix elements with a factor of $1/\sqrt{2}$ for each pair of identical particles in the initial
or final state.
The inverse $T$-matrix is then given by
\beq
T^{-1}_l = A^{-1}_l + G \ ,
\eeq
where the loop integrals $G_{m \bar{m}}$ for the different channels are collected in the diagonal matrix $G$, 
and inversion yields
\beq \label{eq:pwTmat}
T_l = [\mathds{1} + A_l \cdot G]^{-1} \; A_l \ .
\eeq
The expansion of Eq.~(\ref{eq:pwTmat}) in powers of $A_l \cdot G$
\beq \label{eq:BSEex}
T_l = A_l - A_l \cdot G \cdot A_l + \cdots \ .
\eeq
can then be linked to the loopwise expansion of ChPT.
In fact $T_l$ amounts to the summation of a bubble chain to all orders in the $s$-channel,
equivalent to solving a Bethe-Salpeter equation with $A_l$ as driving term,
where all momenta in $A_l$ are set to their on-shell values.

The partial-wave expansion of the $T$-matrix may be cast in Lorentz covariant form. For the scattering of 
particles with four-momenta $q_i$, $\bar{q}_i$ into particles with four-momenta $q_f$, $\bar{q}_f$ we 
define the Mandelstam variables $\hat{s} = (q_i + \bar{q}_i)^2 = p^2$, $\hat{t} = (q_i - q_f)^2$ and 
$\hat{u} = (q_i - \bar{q}_f)^2$. By means of the covariant operators $J_l = J_l(\hat{s}, \hat{t}-\hat{u})$
{\arraycolsep0pt\begin{eqnarray}
J_s=&&\mathrel{}1\,,\nonumber\\
J_p=&&\mathrel{}h_{\mu\nu}q_i^\mu q_f^\nu
=\frac{\hat{t}-\hat{u}}{4}+\frac{(q_i^2-\bar q_i^2)(q_f^2-\bar q_f^2)}{4\hat{s}}\,,
\nonumber\\
J_d=&&
J_p^2-\frac{h_{\mu\nu}q_i^\mu q_i^\nu\,h_{\rho\sigma}q_f^\rho q_f^\sigma}{3} \,,
\end{eqnarray}}%
with
\begin{equation}
h_{\mu\nu}=-g_{\mu\nu}+p_\mu p_\nu/p^2
\end{equation}
the partial-wave expansions of $A$ and $T$ can be rewritten as
\begin{equation}\label{eq:TPartialWave}
\begin{array}{c}
A=\tsum\nolimits_l \hat{A}_l J_l=\hat{A}_s J_s+\hat{A}_p J_p+\hat{A}_d J_d \ ,\\[1.0ex]
T=\tsum\nolimits_l \hat{T}_l J_l=\hat{T}_s J_s+\hat{T}_p J_p+\hat{T}_d J_d \ ,
\end{array}
\end{equation}
where the $\hat{A}_l$, $\hat{T}_l$ only depend on $\hat{s}$.
The $\hat{A}_l$ and $\hat{T}_l$ are related to the original partial waves
$A_l$ and $T_l$ by means of a (diagonal) transformation matrix $C_l$ 
\beq
A_l = C_l \,\hat{A}_l \,C_l, \qquad T_l = C_l \,\hat{T}_l \,C_l,
\eeq
and in analogy to Eq.~(\ref{eq:BSEex}) we can write down a Bethe-Salpeter equation for $\hat{T}_l$
\beq \label{eq:BSE}
\hat{T}_l = \hat{A}_l - \hat{T}_l \,\hat{G}_l \,\hat{A}_l
\eeq
with $\hat{G}_l = C_{l} \,G \,C_{l}$.
In the present work, we restrict ourselves to $s$- and $p$-waves and drop the $d$-wave part $\hat{T}_d$.

We are now in a position to include infinite rescattering processes 
of meson pairs, which are incorporated in $T$,
as final state interactions in the decay amplitudes. In order to introduce a common notation for the decays 
investigated in the present work, we define Mandelstam variables 
\beq \label{eq:Mandelstam}
s = (p_h - p_i)^2, \quad t = (p_h - p_j)^2, \quad u = (p_h - p_k)^2
\eeq
for the generic process $h \to i\,j\,k$ and the $p_x$ represent the momenta of the particles. 
Since all decays under consideration happen to have a 
particle-antiparticle pair in the final state, i.e.\ either $\pi^+ \pi^-$ or $\pi^0 \pi^0$,
which we denote by $j$ and $k$ with $j = \bar{k}$,
$C$-invariance dictates that the decay amplitude $\mathcal{A}_{hijk}(s,t,u)$ is symmetric under 
$t \leftrightarrow u$.
The full amplitude $\mathcal{A}_{hijk}$, which includes $s$- and $p$-wave final state interactions, is constructed
in such a way that it reproduces the tree level result and the unitarity corrections from one-loop ChPT. It reads
\begin{multline} \label{eq:decamp}
\mathcal{A}_{hijk}(s,t,u) = A_{hijk}(s,t,u) 
  + \biggl(\sum_{l=0,1} J_l(s,t-u) \big[\hat{T}_l(s) - \hat{A}_l(s)\big]\biggr)_{hi,jk} \\
  + \biggl(\sum_{l=0,1} J_l(t,u-s) \big[\hat{T}_l(t) - \hat{A}_l(t)\big]\biggr)_{hj,ik} 
  + \biggl(\sum_{l=0,1} J_l(u,s-t) \big[\hat{T}_l(u) - \hat{A}_l(u)\big]\biggr)_{hk,ij} \ ,
\end{multline}
where $A_{hijk}$ is the complete tree level amplitude from ChPT up to fourth chiral order. 
The differences
$\hat{T}_l - \hat{A}_l$ are introduced to avoid double-counting of tree graphs and start contributing 
at the one-loop level. Depending on the subscripts of the parentheses $\hat{T}_l$ and $\hat{A}_l$ 
collect either charged or uncharged channels. For identical particles in the final state ($\pi^0 \pi^0$)
they must be multiplied by a combinatorial factor of $\sqrt{2}$, in order to cancel the 
symmetry factor which had been absorbed into the potentials. 
We note that $\mathcal{A}_{hijk}$ does not entail the full one-loop result from ChPT, since 
tadpoles are not included, but they can be absorbed
by redefining the couplings of the effective Lagrangian \cite{OO}.

\subsection{Unitarity} \label{sec:uni}

\begin{figure}
\begin{center}
$2i \Im \quad$
\parbox{2.2cm}{\includegraphics[height=2.5cm]{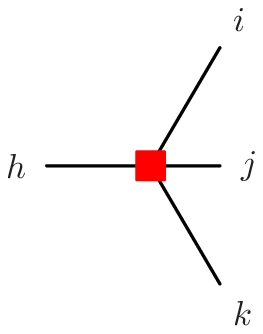}}
$\quad = \quad$
\parbox{3.3cm}{\includegraphics[height=2.5cm]{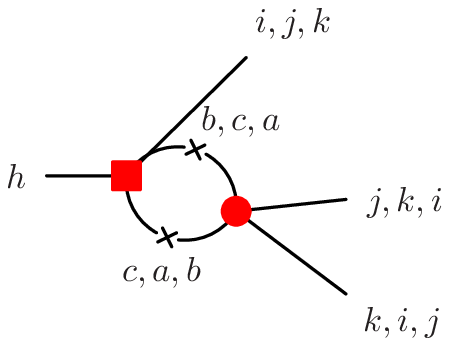}}
$\quad + \quad$
\parbox{3.1cm}{\includegraphics[height=2.5cm]{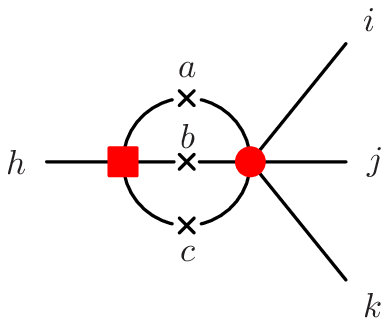}}
\end{center}
\caption{Diagrammatic representation of the unitarity relation in Eq.~(\ref{eq:uni}). The crosses indicate
         on-shell particles and phase space integration.}
\label{fig:uni}
\end{figure}

The approach described above incorporates the relevant pieces of the ChPT one-loop amplitude, fulfills 
unitarity constraints from two-particle scattering and has a perspicuous diagrammatic representation of 
the final state interactions: the summation of a bubble sum in each of the three two-particle channels.
However, it does not account for three-body interactions in the final state, either mediated by the 
interaction of one of the two rescattering particles with the third, spectating particle or by a genuine 
three-body force. Therefore, the approach does not guarantee exact unitarity of the resulting $S$-matrix
which implies a relation for the imaginary part of the decay amplitude $\mathcal{A}_{hijk}$
\cite{Walker}
\beq \label{eq:uni}
\Im{\mathcal{A}_{hijk}} = \frac{1}{2} \sum_{a,b,c} (2\pi)^4 \delta^{(4)}(p_i + p_j + p_k - p_a - p_b - p_c)
                          \ \mathcal{T}^{*}_{abc,ijk} \ \mathcal{A}_{habc} \ ,
\eeq
where $\mathcal{T}^{*}_{abc,ijk}$ represents the complex conjugate of the scattering amplitude for
$ijk \to abc$ and the sum, which includes the integration over phase space, runs over all possible 
three-particle states which $h$ can decay into. A diagrammatic representation of the unitarity condition
is shown in Fig.~\ref{fig:uni}. For $\mathcal{T}_{abc,ijk}$ we make an approximation similar to the one 
already applied to $\mathcal{A}_{hijk}$, i.e., we drop the last diagram on the r.h.s.\ of Fig.~\ref{fig:uni} 
and keep only the graphs involving exclusively two-particle rescattering.
The first term on the r.h.s. can be expressed in terms of the unitarized two-body scattering 
amplitude $T$, and Eq.~(\ref{eq:uni}) reduces to
\beq \label{eq:unired}
\begin{array}{l}
\Im{\mathcal{A}_{hijk}}(s,t,u) \\ 
= \displaystyle\frac{1}{16\pi^2} \sum_{b,c} \int \frac{d^3 p_b}{2 p_{b}^0} \frac{d^3 p_c}{2 p_{c}^0} \sum_{l=0,1}
  \Big\{J_l(s,t'-u')\,[\hat{T}_l^{bc,jk}(s)]^*\,\mathcal{A}_{hibc}(s,t',u')\,\delta^{(4)}(p_b + p_c - p_j - p_k)\Big\} \\
\quad \displaystyle +\frac{1}{16\pi^2} \sum_{a,c} \int \frac{d^3 p_a}{2 p_{a}^0} \frac{d^3 p_c}{2 p_{c}^0} \sum_{l=0,1}
  \Big\{J_l(t,u'-s')\,[\hat{T}_l^{ac,ik}(t)]^*\,\mathcal{A}_{hajc}(s',t,u')\,\delta^{(4)}(p_a + p_c - p_i - p_k)\Big\} \\
\quad \displaystyle +\frac{1}{16\pi^2} \sum_{a,b} \int \frac{d^3 p_a}{2 p_{a}^0} \frac{d^3 p_b}{2 p_{b}^0} \sum_{l=0,1}
  \Big\{J_l(u,s'-t')\,[\hat{T}_l^{ab,ij}(u)]^*\,\mathcal{A}_{habk}(s',t',u)\,\delta^{(4)}(p_a + p_b - p_i - p_j) \Big\},
\end{array}
\eeq
where, in analogy to the definition of $\mathcal{A}$, we only include two-particle rescattering in the $s$ and 
$p$ partial wave. The Mandelstam variables $s'$, $t'$, $u'$ are defined as
\beq
s' = (p_h - p_a)^2\,, \qquad t' = (p_h - p_b)^2\,, \qquad u' = (p_h - p_c)^2\,.
\eeq

\begin{figure}
\begin{center}
\includegraphics[width=5cm]{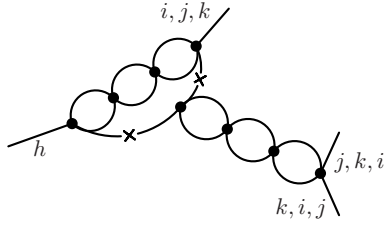}
\end{center}
\caption{Contribution which is included on the r.h.s.\ of Eq.~(\ref{eq:uni}), but not in $\mathcal{A}$ on the
         l.h.s. The crosses indicate on-shell particles and phase space integration.}
\label{fig:3body}
\end{figure}

The two spectator approximations utilized for $\mathcal{A}$ and the r.h.s.\ of Eq.~(\ref{eq:unired}),
however, do not coincide, since contributions like the one shown
in Fig.~\ref{fig:3body} appear on the r.h.s.\ of Eq.~(\ref{eq:unired}), but are missing in $\mathcal{A}$ on the 
l.h.s. The violation of Eq.~(\ref{eq:unired}) gives us the possibility to estimate the importance of this
class of three-body contributions which goes beyond pure two-particle rescattering as embodied in $\mathcal{A}$.
As we will discuss below, it turns out that these deviations are rather small for $\eta \to 3\pi$ and 
$\eta' \to \eta \pi \pi$ where phase space is narrow and dropping the last term in Fig.~\ref{fig:uni} appears to
be a good approximation. 
Assuming that structures involving more complicated iterated two-body interactions or three-body contact terms
yield contributions of comparable (or smaller) size we conclude that our approach approximates the physical 
amplitude reasonably well.
For $\eta' \to 3\pi$, on the other hand, where phase space is about a factor of seven larger than for 
$\eta \to 3\pi$, neglecting the last term in Fig.~\ref{fig:uni} is no longer justified
and hence Eq.~(\ref{eq:unired}) is not suited to be an appropriate estimate of unitarity corrections not 
included in the approach.
A more detailed study on the importance of three-body effects is beyond the scope
of the present investigation, but
will be addressed in forthcoming work \cite{BN}.

\subsection{Isospin decomposition} \label{sec:iso}

In order to study the importance of the various two-particle channels in the final state interactions
and corresponding contributions from well-known resonances like $f_0(980)$, $a_0(980)$ we perform a 
decomposition into isospin channels. Assigning one common mass for all particles of an isospin 
multiplet this can straightforwardly be done for the isospin-conserving decay modes $\eta' \to \eta \pi \pi$.
To this aim, we decompose the interaction kernel of the Bethe-Salpeter equation, Eq.~(\ref{eq:BSE}), into an 
isospin-conserving and an isospin-breaking part, $\hat{A}_{l}^{\textrm{(ic)}}$ and $\hat{A}_{l}^{\textrm{(ib)}}$,
respectively, so that
\beq
\hat{A}_l = O \,\hat{A}_{l}^{\textrm{(ic)}} \,O^T + O \,\hat{A}_{l}^{\textrm{(ib)}} \,O^T \ ,
\eeq
where $O$ represents the orthogonal transformation which transforms from the isospin to the physical basis.
Analogous to the definition of $\hat{T}_{l}$
we can construct the unitarized amplitude $\hat{T}_{l}^{\textrm{(ic)}}$ in the isospin limit
by replacing $\hat{A}_l$ by $\hat{A}_{l}^{\textrm{(ic)}}$ in Eq.~(\ref{eq:BSE}).
After substituting in Eq.~(\ref{eq:decamp}) the pieces of the form $[\hat{T}_l - \hat{A}_l]$ by
$O [\hat{T}_{l}^{\textrm{(ic)}} - \hat{A}_{l}^{\textrm{(ic)}}] O^T$, the influence of the different 
isospin channels may be examined by omitting one specific combination of isospin and angular
momentum quantum numbers.

The situation is slightly more complicated for the isospin-breaking decays of $\eta$ and $\eta'$ into three 
pions. Retaining only one isospin-breaking vertex and inserting it at all possible places in the bubble 
chain, the decay amplitude in terms of isospin channels is found by replacing in Eq.~(\ref{eq:decamp}) 
the pieces of the form $[\hat{T}_l - \hat{A}_l]$ by 
\beq
O [\hat{T}_{l}^{\textrm{(ic)}}\,\hat{G}_l\,\hat{A}_{l}^{\textrm{(ib)}}\,\hat{G}_l\,\hat{T}_{l}^{\textrm{(ic)}} 
  -\hat{T}_{l}^{\textrm{(ic)}}\,\hat{G}_l\,\hat{A}_{l}^{\textrm{(ib)}}
  -\hat{A}_{l}^{\textrm{(ib)}}\,\hat{G}_l\,\hat{T}_{l}^{\textrm{(ic)}}] O^T \ .
\eeq
While the second and third term in the bracket describe the insertion of the isospin-breaking vertex at both
ends of the bubble chain, the first one includes insertions at all intermediate points.

\section{Results}  \label{sec:res}

We now turn to the discussion of the numerical results of the calculation which are obtained from a combined 
analysis of the decay widths, branching ratios, and slope parameters of the considered decays as well as 
phase shifts in meson-meson scattering. The widths of $\eta \to 3 \pi$ and $\eta' \to \eta \pi \pi$ have been 
measured roughly at the 10\,\% precision level,
while for $\eta' \to 3 \pi^0$ the experimental uncertainty is considerably larger and only an 
upper limit exists for $\Gamma(\eta' \to \pi^+ \pi^- \pi^0)$ \cite{pdg}. Moreover, some of these decay widths are 
constrained by the well-measured branching ratios
\beq \label{eq:defBR}
r_1 = \frac{\Gamma(\eta \to 3\pi^0)}{\Gamma(\eta \to \pi^+ \pi^- \pi^0)} \,, \qquad
r_2 = \frac{\Gamma(\eta' \to 3\pi^0)}{\Gamma(\eta' \to \eta \pi^0 \pi^0)} \,.
\eeq

The Dalitz plot distribution of the decay $h \to i\,j\,k$ (with $j = \bar{k}$) is conventionally described in terms 
of the two variables
\begin{eqnarray} \label{eq:dalitz}
x &=&\frac{\sqrt{3}(u-t)}{2 m_h (m_h-m_i-2m_{jk})} \ , \nonumber \\
y &=&  \frac{(m_i+2m_{jk})\big[ (m_h-m_i)^2 -s \big]}{2 m_h m_{jk}(m_h-m_i-2m_{jk})} - 1 \ ,
\end{eqnarray}
where the $m_x$ denote the masses of the respective particles ($m_j = m_k = m_{jk}$) and the Mandelstam variables
have been defined in Eq.~(\ref{eq:Mandelstam}). In $\eta \to 3 \pi$ measurements (e.g.\ \cite{Abele}) 
a slightly simpler definition of $y$, where $(m_{\pi^0}+2m_{\pi^+})/m_{\pi^+}$ is replaced by 3, is usually 
employed,
\beq
y = \frac{3 \big[(m_h-m_i)^2 -s \big]}{2 m_h (m_h-m_i-2m_{jk})} - 1 \ ,
\eeq
but the difference is at the level of 1\,\% and can be safely neglected.
The squared absolute value of the amplitude, 
$|\mathcal{A}_{hijk}(x,y)|^2$, is then expanded for $\eta' \to \eta \pi \pi$ and the charged decay modes
of $\eta, \eta' \to 3\pi$ as
\begin{equation} \label{eq:DalC}
|\mathcal{A}(x,y)|^2  = |N|^2 \big[ 1 + a y + b y^2 + c x^2 + d y^3 + \cdots \big]\, ,
\end{equation}
while for the decays into three identical particles Bose symmetry dictates the form
\begin{equation} \label{eq:Dal0}
|\mathcal{A}(x,y)|^2  = |N'|^2 \big[ 1 + g( y^2 + x^2) + \cdots \big] \,.
\end{equation}
For the Dalitz plot parameters $a$, $b$, $c$ of $\eta \to \pi^+ \pi^- \pi^0$ the experimental situation 
is not without controversy. We employ the numbers of \cite{Abele}, since it is the most recent published 
measurement and the results appear to be consistent with the bulk of the other experiments listed by the 
Particle Data Group \cite{pdg}. They differ somewhat from 
the new preliminary results of the KLOE Collaboration \cite{KLOE} that has found a non-zero value for the 
third order parameter $d$ which was not included in previous experimental parametrizations.
Very recently the Dalitz plot parameters 
of $\eta' \to \eta \pi^+ \pi^-$ have been determined with high statistics by the VES experiment \cite{VES}.
While we will employ in our fits the experimental Dalitz parameters provided by the Particle Data Group
\cite{pdg}, we will also discuss further below the modifications of our results when the two new and precise 
(but so far preliminary) data sets \cite{KLOE, VES} are taken into account.
Note that the slope parameters of $\eta' \to 3 \pi$ have not yet been determined experimentally, but such a 
measurement is intended at WASA@COSY \cite{WASA}.

{From} the unitarized partial-wave $T$-matrix in Eq.~(\ref{eq:pwTmat}) one may also derive the phase shifts 
in meson-meson scattering. Hence, our approach is further constrained by the experimental phase shifts for 
$\pi \pi \to \pi \pi, K \bar{K}$ scattering. The results of the fit are presented in Appendix~\ref{sec:AppPhSh}.

The coupled-channels framework entails several parameters, i.e.\ the low-energy constants (LECs) of the chiral 
effective Lagrangian up to fourth order and the subtraction constants $a_{m \bar{m}}$ in the loop integrals 
$G_{m \bar{m}}$ which are embodied in the coupled-channels $T$-matrix. 
It turns out that only the fit to the 
$\pi \pi$ phase shifts in the $I = J = 1$ channel (which includes the $\rho$ resonance) requires a non-zero 
value of the corresponding subtraction constant $a_{\pi \pi}^{(I=J=1)}(\mu)$. 
The regularization scale of $G_{m \bar{m}}$ is set to $\mu = 1$~GeV for all channels.
As a guiding principle for the importance of the LECs 
and in order to reduce their number, we make use of large-$N_c$ arguments in the effective Lagrangian and set 
all LECs to zero which are
of order $\mathcal{O}(1/N_c^2)$ and thus suppressed by at least three powers of $1/N_c$ with 
respect to the leading coefficients. Their effects are expected to be small and can be partially compensated by 
readjusting the leading and subleading coefficients in our fits. Furthermore, we set those parameters to zero 
by hand which turn out to be less sensitive to the processes under consideration.
It turns out, that with the exception of $\coeffv{1}{2}$ and $\cbeta{14}{0}$ all parameters of order 
$\mathcal{O}(1/N_c)$ have a negligible effect when varying them within small ranges around zero
and can be safely neglected. To summarize, we only keep the LECs
\beq
\begin{array}{c}
\cbeta{0}{0},\, \cbeta{3}{0},\, \cbeta{5}{0},\, \cbeta{8}{0} = \mathcal{O}(N_c) \,, \\[1.5ex]
\coeffv{0}{2},\, \coeffv{3}{1},\, \cbeta{1}{0},\, \cbeta{2}{0},\, \cbeta{4}{0},\, \cbeta{6}{0},\, \cbeta{7}{0},\, 
\cbeta{13}{0},\, \cbeta{18}{0},\, = \mathcal{O}(1) \,, \\[1.5ex]
\coeffv{1}{2},\, \cbeta{14}{0} = \mathcal{O}(1/N_c)\,.
\end{array}
\eeq
The coefficient $\coeffv{0}{2}$ is related to the mass of the $\eta'$ in the chiral limit, $m_0$, and has been
constrained to the range $0.00183 \GeV^4 \dots 0.00523 \GeV^4$ in \cite{BB3}, while the rest of the parameters 
of order $\mathcal{O}(N_c^i)$ may be varied within small ranges around zero, naturally given by 
$\pm  N_c^i f^2/(12 \Lambda_{\chi}^n)$, where $n$ depends on the dimension of the constant under consideration.
In conventional ChPT the $\beta_0$ term is usually not listed, since it can be absorbed into other contact 
terms by virtue of a Cayley-Hamilton matrix identity.  However, this transformation
mixes different orders in $1/N_c$, hence we prefer to keep this term explicitly, in order to retain the 
clean large $N_c$ behavior of the $\beta_i$'s.

By fitting to all available (published) data sets of the investigated hadronic $\eta, \eta'$ decays
and the phase shifts 
an overall $\chi^2$ function is calculated. To this end, we compute $\chi^2$ values for all observables i.e.\ 
phase shifts, decay widths, branching ratios and Dalitz plot parametrizations, divide them by the number 
of experimental data points and take the sum afterwards.
In order to find the minima of the overall $\chi^2$ function, we perform
a random walk in parameter space, where only steps which lead to a smaller $\chi^2$ value are allowed, 
and a very large number of random walks with randomized starting points is carried out.
We observe four different classes of fits which are all in very good agreement with the 
currently available (published) data on hadronic decays, but differ in the description of the decays
$\eta' \to 3\pi$ where experimental constraints are scarce.

The errors which we specify in the following for all parameters and observables reflect the deviations 
which arise when we allow for $\chi^2$ values which are at most 15\,\% larger than the minimum value.
Although this choice is somewhat arbitrary, it illustrates how variation 
of the $\chi^2$ function in parameter space affects the results.

\subsection{\mbox{\boldmath $\eta \to 3\pi$}} \label{sec:83xx}

The results for the decays $\eta \to 3\pi$, which agree very well with the experimental values, are shown in 
Table~\ref{tab:eta3pi}. Most remarkably our approach is able to 
reproduce the new, precise value of the $\eta \to 3 \pi^0$ Dalitz 
parameter $g$ measured by the Crystal Ball Collaboration \cite{Tippens} (and prevailing the PDG average 
value \cite{pdg}) which could not be met in previous investigations \cite{Kambor, BB1}. 
With regard to \cite{BB1} this is mainly due to 
the larger number of chiral parameters taken into account in the present work
and the improved fitting routine utilized.
The PDG number for $g$ does, however,
not agree with the preliminary $g$ value of the KLOE Collaboration \cite{KLOE} which comes close to the 
results given in \cite{Kambor, BB1}. The detailed discussion of the new KLOE results for the Dalitz plot 
distributions of $\eta \to \pi^+ \pi^- \pi^0$ and $\eta \to 3\pi^0$ is deferred to Secs.~\ref{sec:newexp}
and \ref{sec:DalPhe}.

When electromagnetic effects are neglected (which is justified according to Sutherland's theorem 
\cite{Sutherland:1966mi}), the isospin-violating decay of $\eta$ into three pions can only take place via 
a finite quark mass difference $m_u - m_d$. The decay amplitude is therefore proportional to $m_{\epsilon}^2$ 
defined in Eq.~(\ref{eq:meps}) and we have employed the value which follows from Dashen's theorem, 
Eq.~(\ref{eq:Dashen}) \cite{Dashen}. Deviations of the calculated decay widths from the measured numbers 
could thus be interpreted as a hint to non-negligible subleading corrections to the leading order result 
by Dashen. In order to quantify these deviations, one commonly defines the double quark mass ratio
\begin{equation}
Q^2 = \frac{m_s - \hat{m}}{m_d - m_u} \frac{m_s + \hat{m}}{m_d + m_u} \ ,
\end{equation}
and Dashen's theorem yields $Q_{\textrm{Dashen}} = 24.1$. Differing $Q$-values lead 
to decay widths which are related to the original one, $\Gamma_{\textrm{Dashen}}$, by 
\beq
\Gamma = \left(\frac{Q_{\textrm{Dashen}}}{Q}\right)^4 \Gamma_{\textrm{Dashen}} \ .
\eeq
Taking into account theoretical as well as experimental uncertainties 
we find from a comparison of our results with data $Q = 24.0 \pm 0.6$
which is consistent with the result of \cite{BB1}.
Note, however, that this obvious agreement with Dashen's theorem merely reflects the fact that our 
approach is capable of reproducing the experimental decay widths of $\eta \to 3\pi$. 
Due to the larger number of chiral parameters with increased ranges compared to \cite{BB1}
and the improved fitting procedure
we can easily compensate the effects from variations in $Q$ 
by readjusting the chiral parameters of our approach.
We have checked that variations of $Q$ in the 
range of $20 \dots 24$ which covers 
the various (and partially contradictory) results in the literature \cite{CorrDash}
can be accommodated within this approach.
Therefore, our analysis does not allow for conclusions on the size of the violation of Dashen's theorem.

Extending the work of \cite{BB1} we have also taken $p$-wave final state interactions into account. By setting 
these contributions to zero, we find that the decay width of $\eta \to \pi^+ \pi^- \pi^0$ is reduced by a tiny
fraction of 0.7\,\% implying rapid convergence of the partial wave expansion. 
The Dalitz plot parameters, which are more sensitive to the precise form of the amplitude than the width,
are also only moderately altered. Without $p$-waves we obtain $a = -1.15 \pm 0.07$, $b = 0.29 \pm 0.05$, 
$c = 0.01 \pm 0.02$.
Note that due to Bose symmetry there is no $p$-wave contribution to the decay into three neutral pions.

Certainly, the most important isospin channel for final state interactions in $\eta \to 3 \pi$ is the
$I=0$ $s$-wave rescattering which is dominated by $\pi \pi$ interactions. Omitting this channel reduces the 
decay width by 73\,\%. The other two $s$-wave channels with isospin one and two, respectively, interfere destructively with 
the former. To be more precise, taking out the $I=1$ part, which mainly reflects $\pi \eta$ interactions, enlarges the decay width 
of $\eta \to \pi^+ \pi^- \pi^0$ ($\eta \to 3 \pi^0$) by 9\,\% (10\,\%), while setting the $I=2$ channel, 
which is purely $\pi \pi$ rescattering, to zero results in an enhancement of the decay widths by 16\,\% (20\,\%).
The only relevant $p$-wave contributions arise from the $I=1$ channels $\pi \eta$, $\pi \eta'$ which are
$C$-even and thus do not couple to $C$-odd channels related to the $\rho(770)$ resonance.
Neglecting the $I = J = 1$ channels reduces the 
$\eta \to \pi^+ \pi^- \pi^0$ decay width by roughly 1\,\%. The numerical difference to the statement on the 
importance of $p$-wave contributions in the previous paragraph is due to the fact that for the decomposition 
into isospin channels we use isospin-symmetrized masses, while otherwise we employ the physical values of the 
masses.

Finally, we have also examined to which extent the amplitude violates three-particle unitarity in the spectator
approximation as described in Sec.~\ref{sec:uni}. In order to quantify the violation of Eq.~(\ref{eq:unired}),
we compute the absolute value of the difference between l.h.s.\ and r.h.s.\ normalized by the modulus of the
amplitude $\mathcal{A}_{hijk}$. Averaged over the whole Dalitz plot we find this violation to be
$(2.5 \pm 0.3)\,\%$ for the process $\eta \to \pi^+ \pi^- \pi^0$ and---even smaller---$(1.3 \pm 0.3)\,\%$
for the decay into three neutral pions. The fact that the violation of Eq.~(\ref{eq:unired}) is so small is
non-trivial since only two-body unitarity, but not three-body unitarity is implemented in the definition of 
the decay amplitude. This may suggest that three-body effects (like multiple scattering of one particle in 
the final state with the other two or a genuine three-body interaction) are of the same order of magnitude.
A more detailed investigation of this issue will be the subject of future work \cite{BN}.

\begin{table}
\centering
\begin{tabular}{|c|D{+}{\,\pm\,}{5,5}|D{+}{\,\pm\,}{4,4}|D{+}{\,\pm\,}{4,4}|D{+}{\,\pm\,}{6,6}|}
\hhline{----}
& \multicolumn{1}{c|}{$\Gamma_{\eta \to 3\pi^0}$ (eV)} & 
\multicolumn{1}{c|}{$\Gamma_{\eta \to \pi^+ \pi^- \pi^0}$ (eV)} &
\multicolumn{1}{c|}{$r_1$} \\
\hhline{----}
theo. &
  422    +    13    &
  290    +     8    &
  1.46   +   0.02   \\
\hhline{----}
exp. &
  419    +    27    &
  292    +    21    &
  1.44   +   0.04   \\
\hhline{====-}
& \multicolumn{1}{c|}{$a$} &
\multicolumn{1}{c|}{$b$} &
\multicolumn{1}{c|}{$c$} &
\multicolumn{1}{c|}{$g$} \\
\hline
theo. &
 -1.20   +   0.07   &
  0.28   +   0.05   &
  0.05   +   0.02   &
 -0.062  +   0.006  \\
\hline
exp. &
 -1.22   +   0.07   &
  0.22   +   0.11   &
  0.06   +   0.02   &
 -0.062  +   0.008  \\
\hline
\end{tabular}
\caption{Results for the partial decay widths of $\eta \to 3\pi$, the branching ratio $r_1$, and
         the Dalitz plot parameters compared to experimental data from \cite{pdg} and \cite{Abele}.}
\label{tab:eta3pi}
\end{table}

\subsection{\mbox{\boldmath $\eta' \to 3\pi$}} \label{sec:93xx}

\begin{figure}
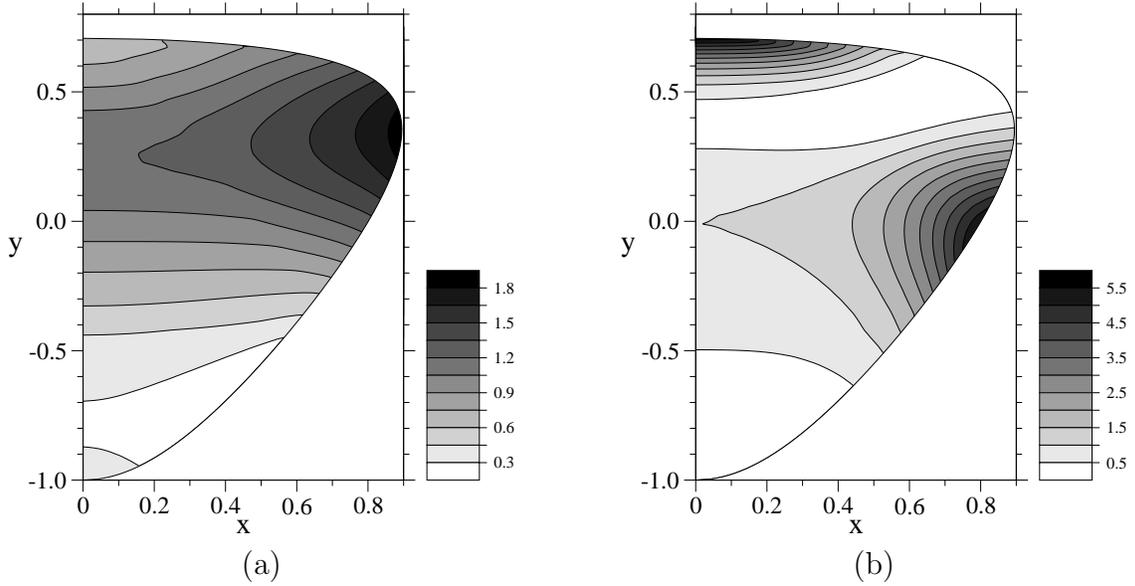

\centering
\begin{tabular}{cp{0.5cm}c}
\includegraphics[width=0.4\textwidth]{cntr9312Cl2.eps} & &
\includegraphics[width=0.4\textwidth]{cntr9312Cl3.eps} \\
(a) & & (b)
\end{tabular}
\caption{Sample $\eta' \to \pi^+ \pi^- \pi^0$ Dalitz plot distribution 
         $|\mathcal{A}(x,y)/\mathcal{A}(0,0)|^2$ of cluster~2 which can be described
         by a fourth order polynomial in $x$ and $y$ (a) and of cluster~3 which must be parametrized
         by a polynomial of eighth order (b). Due to their symmetry under $x \to -x$ only the right half 
         of the Dalitz plots is shown.}
\label{fig:Daletap3pi}
\end{figure}

Only sparse experimental information exists on the decays of $\eta'$ into three pions. 
The experimental decay width of $\eta' \to 3 \pi^0$ is \cite{pdg}
\beq
\Gamma^{\textrm{(exp)}}(\eta' \to 3 \pi^0) = (315 \pm 78) \eV
\eeq
which is nicely met within our approach:
\beq
\Gamma^{\textrm{(theo)}}(\eta' \to 3 \pi^0) = (330 \pm 33) \eV \,.
\eeq
For the decay into $\pi^+ \pi^- \pi^0$ only a weak experimental upper limit exists \cite{pdg}
\beq
\Gamma^{\textrm{(exp)}}(\eta' \to \pi^+ \pi^- \pi^0) < 10 \keV \,.
\eeq
Due to the large phase space available in these two decay modes of the $\eta'$, final state
interactions are expected to be of greater importance. Indeed we find that in contrast to the processes
$\eta \to 3\pi$ and $\eta' \to \eta \pi \pi$ the Dalitz plot distribution of $\eta' \to 3\pi$---depending
on the choice of the chiral parameters---cannot 
always be well parametrized by a simple second or third order polynomial in $x$ and $y$.
Nevertheless, it happens that all our fits may be classified into four groups mainly due to the 
different values of the lower order coefficients in $x$ and $y$. 
The numerical results for these most relevant coefficients 
are compiled in Tables~\ref{tab:9312} and \ref{tab:9333} along with the predicted width of 
$\eta' \to \pi^+ \pi^- \pi^0$ and the order of the polynomial in $x$ and $y$ which 
is needed to obtain a reasonable approximation to the Dalitz plot distribution resulting from our
approach. Note that due to charge conjugation invariance, only even powers of $x$ appear. 
Examples of two very different Dalitz plots are shown in Fig.~\ref{fig:Daletap3pi}. Despite these differing
predictions one should keep in mind that all fits describe all {\it available} experimental data at the same 
level of accuracy.
The Dalitz plot distributions of these decays pose therefore tight constraints for our approach and 
must be compared with future experiments at the WASA@COSY facility.

\begin{table}
\centering
\begin{tabular}{|lll|D{+}{\,\pm\,}{5,5}|D{+}{\,\pm\,}{5,5}|D{+}{\,\pm\,}{5,5}|D{+}{\,\pm\,}{5,5}|}
\hline
& & & \multicolumn{1}{c|}{cluster 1} & \multicolumn{1}{c|}{cluster 2} & \multicolumn{1}{c|}{cluster 3}
& \multicolumn{1}{c|}{cluster 4} \\
\hline
\multicolumn{3}{|l|}{$\Gamma_{\eta' \to \pi^+ \pi^- \pi^0}$ (eV)} 
                                  & 470  + 200  & 520  + 200  & 740  + 420  & 620  + 180  \\
\hline
coeff. & $y$       & (``$a$'')    &  0.6 +  5.2 &  2.4 +  1.7 &  0.3 +  1.1 &  4.4 +  1.2 \\
\hline
coeff. & $y^2$     & (``$b$'')    & 10.0 + 11.0 &  2.1 +  7.5 & -5.2 +  1.5 & 14.9 +  6.7 \\
\hline
coeff. & $x^2$     & (``$c$'')    &  0.1 + 3.6  & -0.7 +  1.4 &  0.1 +  1.6 & -3.7 +  1.5 \\
\hline
coeff. & $y^3$     & (``$d$'')    & -6.1 + 11.5 & -0.6 + 14.0 & -8.8 +  7.8 & 27.5 + 18.1 \\
\hline
coeff. & $x^2 y$   &              &-10.8 + 11.2 &  2.0 +  3.0 & -7.4 +  5.6 & -1.5 +  2.8 \\
\hline
coeff. & $y^4$     &              &  0.6 + 12.2 & -3.2 +  7.3 & 23.3 + 20.7 & 24.5 + 11.6 \\
\hline
coeff. & $x^2 y^2$ &              & 13.9 + 23.6 & 11.8 + 22.4 &-17.7 +  9.4 & 39.0 + 12.7 \\
\hline
coeff. & $x^4$     &              & -0.5 + 11.5 & -1.2 + 16.1 & 15.4 +  9.8 &-20.5 +  9.4 \\
\hline
\multicolumn{3}{|l|}{poly.\ order} & \multicolumn{1}{c|}{6 -- 8} & \multicolumn{1}{c|}{4 -- 8} 
& \multicolumn{1}{c|}{$\geq 8$} & \multicolumn{1}{c|}{$\geq 8$} \\
\hline
\end{tabular}
\caption{Results for the decay width of $\eta' \to \pi^+ \pi^- \pi^0$ and the leading Dalitz plot
         parameters. The last line denotes the order of the polynomial which is needed to describe
         the Dalitz plot distribution.}
\label{tab:9312}
\end{table}

\begin{table}
\centering
\begin{tabular}{|lll|D{+}{\,\pm\,}{5,5}|D{+}{\,\pm\,}{5,5}|D{+}{\,\pm\,}{5,5}|D{+}{\,\pm\,}{5,5}|}
\hline
& & & \multicolumn{1}{c|}{cluster 1} & \multicolumn{1}{c|}{cluster 2} & \multicolumn{1}{c|}{cluster 3}
& \multicolumn{1}{c|}{cluster 4} \\
\hline
coeff. & $x^2$, $y^2$ & (``$g$'') &  0.1 + 1.7 & -2.7 + 1.0 & -2.1 + 0.7 & -0.2 + 0.6 \\
\hline
coeff. & $y^3$        &           & -0.5 + 1.4 & -1.7 + 0.7 & -0.2 + 0.6 & -0.8 + 0.6 \\
\hline
coeff. & $x^2 y$      &           &  1.6 + 4.1 &  5.0 + 1.9 &  0.6 + 1.8 &  2.3 + 1.7 \\
\hline
coeff. & $y^4$        &           &  0.2 + 1.4 &  2.6 + 1.5 &  1.6 + 0.8 & -0.1 + 1.1 \\
\hline
coeff. & $x^2 y^2$    &           &  0.4 + 2.9 &  5.3 + 2.8 &  3.5 + 1.7 &  0.2 + 2.5 \\
\hline
coeff. & $x^4$        &           &  0.1 + 1.5 &  2.7 + 1.5 &  1.7 + 0.9 &  0.1 + 1.2 \\
\hline
\multicolumn{3}{|l|}{poly.\ order} & \multicolumn{1}{c|}{3 -- 6} & \multicolumn{1}{c|}{5 -- 6} 
& \multicolumn{1}{c|}{4 -- 6} & \multicolumn{1}{c|}{3 -- 5} \\
\hline
\end{tabular}
\caption{Results for the leading Dalitz plot parameters of $\eta' \to 3\pi^0$.
         The last line denotes the order of the polynomial which is needed to describe
         the Dalitz plot distribution.}
\label{tab:9333}
\end{table}

While in $\eta' \to 3 \pi^0$ $p$-wave contributions in two-body rescattering are forbidden by Bose symmetry, 
they can be large in $\eta' \to \pi^+ \pi^- \pi^0$ due to large phase space. Interestingly, their size
varies significantly depending on the cluster of fit parameters. They are largest for the fits of 
cluster~4 where setting them to zero diminishes the decay width by 50\,\% on average. The partial width
is reduced by 44\,\% (28\,\%) for cluster~3 (cluster~1), while for the parameter sets of cluster~2 suppressing 
the $p$-wave contributions alters the width by less than 10\,\%.
The large higher order coefficients of the $\eta' \to \pi^+ \pi^- \pi^0$ Dalitz plot distribution 
are mainly due to $p$-wave contributions. If $p$-waves are omitted,
the fits of clusters~3 and 4 can be well parametrized by polynomials
of fifth order in $x$ and $y$, while for most fits in clusters~1
and 2 a sixth order polynomial would be sufficient, cf.\ Table~\ref{tab:9312}.
Note that in analogy to the decay $\eta \to 3 \pi$ $p$-wave final state interactions with
the quantum numbers of the $\rho(770)$ meson do not occur.

In $\eta' \to 3 \pi$ the contributions from the various isospin channels depend sensitively on the cluster,
e.g., omitting the $I = J = 0$ channel in $\eta' \to 3 \pi^0$ reduces the decay width by 84\,\% for cluster~1,
while it is enhanced by 132\,\% on average for the fits of cluster~3. For brevity we refrain from giving the 
full list of isospin contributions.

\subsection{\mbox{\boldmath $\eta' \to \eta \pi \pi$}} \label{sec:98xx}

In Tables~\ref{tab:9812} and~\ref{tab:9833} we show the results for the dominant hadronic decay modes of 
the $\eta'$, namely the decays into $\eta \pi^+ \pi^-$ and $\eta \pi^0 \pi^0$. They are all in very 
good agreement with the existing (and published) experimental data. Furthermore, we have calculated the 
branching ratio $r_2$, Eq.~(\ref{eq:defBR}), which links the two neutral decay modes $\eta' \to 3 \pi^0$ and 
$\eta' \to \eta \pi^0 \pi^0$. We find
\beq
r_{2}^{\textrm{(theo)}} = (71 \pm  7)\times 10^{-4}, \qquad 
r_{2}^{\textrm{(exp)}}  = (74 \pm 12)\times 10^{-4}  \quad \mbox{\cite{pdg}},
\eeq
and the accordance with experiment is again persuasive.

In the isospin limit, $m_u = m_d$, the decay width $\Gamma(\eta' \to \eta \pi^+ \pi^-)$ would be exactly given by 
2 $\Gamma(\eta' \to \eta \pi^0 \pi^0)$, due to the symmetry factor for identical particles in the latter process. 
If, however, one is interested in the isospin-breaking contributions in the amplitude of $\eta' \to \eta \pi \pi$,
one ought to disentangle it from phase space effects which are caused by the different masses of charged and neutral
pions. With an isospin-symmetric decay amplitude, but physical masses in the phase space factors, we find a ratio
\beq
r_3 = \frac{\Gamma(\eta' \to \eta \pi^+ \pi^-)}{\Gamma(\eta' \to \eta \pi^0 \pi^0)} = 1.78 \pm 0.02 \,,
\eeq
which is smaller than 2 and compares to $r_3 = 1.77 \pm 0.02$
when isospin-breaking is taken into account in the amplitude.
(For comparison, if the amplitude is set constant and the physical pion masses are employed in the phase 
space integrals, the ratio is given by $r_3 = 1.77$.)
We may thus conclude that within our approach 
isospin-breaking corrections in the $\eta' \to \eta \pi \pi$ decay amplitude are tiny.
The branching ratio $r_3$ has not been measured directly. If, however, we calculate the ratio of fractions
$\Gamma_i / \Gamma_{\textrm{total}}$ for these two decay modes using the numbers and correlation coefficients
published by the Particle Data Group \cite{pdg}, we arrive at
\beq
r_{3}^{\textrm{(exp)}} = 2.12 \pm 0.19
\eeq
by means of standard error propagation.
Such a large branching ratio would indicate significant isospin-violating contributions
in the amplitude. But the experimental uncertainties are sizable
and should be reduced by the upcoming experiments with WASA at COSY \cite{WASA} and at MAMI-C \cite{Nef}.

\begin{table}
\centering
\begin{tabular}{|c|r@{$\,\pm\,$}l@{\ }l|r@{$\,\pm\,$}l|r@{$\,\pm\,$}l|r@{$\,\pm\,$}l|}
\hline
& \multicolumn{3}{c|}{$\Gamma_{\eta' \to \eta \pi^+ \pi^-}$} &
\multicolumn{2}{c|}{$a$} &
\multicolumn{2}{c|}{$b$} &
\multicolumn{2}{c|}{$c$} \\
\hline
theo. &
$ 81   $ & $  4   $ & keV &
$-0.116$ & $ 0.024$ &
$ 0.000$ & $ 0.019$ &
$ 0.016$ & $ 0.035$ \\
\hline
exp. &
$ 89   $ & $ 11   $ & keV &
$-0.16 $ & $ 0.06 $ &
\multicolumn{2}{l|}{ } & 
\multicolumn{2}{l|}{ } \\
\hline
\end{tabular}
\caption{Results for the partial decay width of $\eta' \to \eta \pi^+ \pi^-$ and the Dalitz plot 
         parameters compared to experimental data from \cite{pdg}.}
\label{tab:9812}
\end{table}

\begin{table}
\centering
\begin{tabular}{|c|r@{$\,\pm\,$}l@{\ }l|r@{$\,\pm\,$}l|r@{$\,\pm\,$}l|r@{$\,\pm\,$}l|}
\hline
& \multicolumn{3}{c|}{$\Gamma_{\eta' \to \eta \pi^0 \pi^0}$} &
\multicolumn{2}{c|}{$a$} &
\multicolumn{2}{c|}{$b$} &
\multicolumn{2}{c|}{$c$} \\
\hline
theo. &
$ 46   $ & $  3   $ & keV &
$-0.122$ & $ 0.025$ &
$ 0.003$ & $ 0.018$ &
$ 0.019$ & $ 0.039$ \\
\hline
exp. &
$ 42   $ & $  6   $ & keV &
$-0.116$ & $ 0.026$ &
$ 0.003$ & $ 0.017$ &
$ 0.00 $ & $ 0.03 $ \\
\hline
\end{tabular}
\caption{Results for the partial decay width of $\eta' \to \eta \pi^0 \pi^0$ and the Dalitz plot 
         parameters compared to experimental data from \cite{pdg}.}
\label{tab:9833}
\end{table}

It turns out that $p$-wave final state interactions are tiny in the processes $\eta' \to \eta \pi \pi$.
The corrections to the decay widths which they generate are smaller than 0.02\,\% and can thus be safely 
neglected.
Consequently, in the isospin basis the relevant two-body channels are given by $s$-wave interactions of
isospin 0 or 1 states. When examining the influence of these two channels on the $\eta' \to \eta \pi \pi$
partial widths we observe an interesting pattern. By setting the $I=0$ channel to zero for the fits of 
cluster~1 (cluster~2) the widths are lowered by 22\,\% (22\,\%), while suppressing the $I=1$ part reduces them 
by 81\,\% (72\,\%). When the fit parameters of clusters~1 and 2 are employed, the isospin one channel which 
includes the tail of the $a_0(980)$ resonance thus appears to be of great importance for the decay mode
$\eta' \to \eta \pi \pi$ confirming the findings of \cite{BB1} and \cite{Fariborz}.
The situation is, however, reversed if one considers the fits of the remaining two clusters. Taking out the 
$I=0$ channel in the final state interactions of the fits of cluster~3 (cluster~4) diminishes the decay 
widths by 79\,\% (81\,\%), whereas erasing the channel with $I=1$ reduces it by only 33\,\% (28\,\%).
Accordingly, for these sets of parameters the $I=0$ channel, which incorporates the effects of the 
$f_0(980)$ resonance and the $\pi \pi$ correlation at lower energies, 
has higher impact on the decay widths than the $a_0(980)$ channel.
Although the fits of all four clusters yield very similar results for all $\eta' \to \eta \pi \pi$
observables, the two scenarios can be distinguished by their correlation with the processes
$\eta' \to 3 \pi$ provided within our approach. Thus, a precise measurement of $\eta' \to 3 \pi$ decay 
parameters can also help to clarify the importance of $a_0(980)$ or $f_0(980)$ resonance contributions 
to the dominant decay mode of the $\eta'$ into $\eta \pi \pi$.

The violation of three-particle unitarity as described in Sec.~\ref{sec:uni} is not as tiny as in the case 
of $\eta \to 3 \pi$, but still remarkably small. Using the definition of Sec.~\ref{sec:83xx} we find averaged 
deviations of $(11 \pm 7)\,\%$ for $\eta' \to \eta \pi^+ \pi^-$ and $(10 \pm 6)\,\%$ for 
$\eta' \to \eta \pi^0 \pi^0$. It remains to be seen whether corrections from other three-body effects 
which are not included in the approach will be of comparable size \cite{BN}.

\subsection{Numerical values of the chiral parameters and \mbox{\boldmath $\eta$}-\mbox{\boldmath $\eta'$} mixing}

Before presenting numerical results for the chiral parameters
we would like to stress  that the values of the couplings of the
effective Lagrangian employed in the coupled-channels approach are in general not identical
to those in the perturbative framework. 
First, contributions from tadpoles (which include also effects from the
so-called on-shell approximation), and $t$-/$u$-channel diagrams in the interaction kernel
have been absorbed into the coupling constants.
Second, the BSE summarizes meson-meson scattering in the $s$-channel to infinite order.
The contributions beyond a given chiral order are missing in the perturbative approach and
lead to changes in the values of the couplings when fitting the results to data.
Finally, the subtraction point in the renormalization procedure can be different
in both schemes.
Hence, one must expect differences in the values of the coupling constants utilized
in both frameworks.

In Table~\ref{tab:LECs} we show the numerical values of the low-energy constants as well as the non-zero
subtraction constant $a_{\pi \pi}^{(I=J=1)}$ as they come out for the fits of the four different clusters.
In addition we display the parameters of $\eta$-$\eta'$ mixing $R_{0 \eta}$, $R_{8 \eta'}$ which 
are determined by the values of the LECs $\coeffv{3}{1}$, $\cbeta{5}{0}$, and $\cbeta{18}{0}$ in virtue of
Eq.~(\ref{eq:mixpar}).
Note that compared to the analysis in \cite{BB1} we have increased the number of chiral parameters 
which---in conjunction with an improved fitting procedure---helped to considerably improve the agreement 
with experimental data on hadronic $\eta$, $\eta'$ decays.

According to the mixing parameters the four clusters of fits may be divided into two groups. For clusters~1
and 2 $R_{0 \eta}$ and $R_{8 \eta'}$ are both of similar small size and (mainly) positive, while clusters~3
and 4 feature a large, positive $R_{0 \eta}$ and an $R_{8 \eta'}$ which is close to zero.
Within the present analysis the second mixing parameter $R_{8 \eta'}$, which 
characterizes the fraction of the pure octet field $\eta_8$ in the physical $\eta'$, turns out to be 
more tightly constrained by the fit than $R_{0 \eta}$ which describes the 
singlet content of the $\eta$.
In all cases the numerical results for $R_{0 \eta}$ and $R_{8 \eta'}$ deviate sizably from an orthogonal 
mixing scheme, where $R_{0 \eta} = -R_{8 \eta'}$. For comparison, a mixing angle of $-20^\circ$ 
in the one-mixing angle scheme as found in 
the literature \cite{GL} would correspond to $R_{0 \eta} = -R_{8 \eta'} = 0.34$.

The fitting procedure does not constrain all parameters at the same level of accuracy. While some 
(e.g.\ $\coeffv{3}{1}$, $\cbeta{3}{0}$, $\cbeta{5}{0}$, $\cbeta{8}{0}$, $\cbeta{18}{0}$) may vary within 
large ranges (partly compensating each other), others like $\coeffv{0}{2}$, $\cbeta{0}{0}$, $\cbeta{1}{0}$, 
$\cbeta{2}{0}$, and $\cbeta{4}{0}$ are relatively tightly fixed. These boundaries constitute important 
constraints which must be met in future coupled-channels analyses of mesonic processes within the 
approach described here.
In particular, the coefficient $\coeffv{0}{2}$ encodes the mass of the $\eta'$ in the chiral limit, 
$m_0$, by virtue of
\beq
m_{0}^{2} = \frac{2 \coeffv{0}{2}}{f^2} \,.
\eeq
The fact that the $\eta'$ does not become massless in the chiral limit is a consequence of the axial
U(1) anomaly of QCD which generates in the divergence of the singlet axial-vector current 
an additional, non-vanishing term involving the gluonic field strength
tensor. In the effective theory this term is represented by $\coeffv{0}{2}$.
Employing $f = 88 \MeV$, the value of the pseudoscalar decay constant in the chiral limit \cite{GL3},
we find from the fits of all clusters $m_0 = (900 \pm 80) \MeV$ which is close to the physical mass
of the $\eta'$.

\begin{table}
\centering
\begin{tabular}{|r@{}l|D{+}{\,\pm\,}{5,5}|D{+}{\,\pm\,}{5,5}|D{+}{\,\pm\,}{5,5}|D{+}{\,\pm\,}{5,5}|}
\hline
& & \multicolumn{1}{c|}{cluster 1} & \multicolumn{1}{c|}{cluster 2} & \multicolumn{1}{c|}{cluster 3}
& \multicolumn{1}{c|}{cluster 4} \\
\hline
$\coeffv{3}{1}$ & $\times 10^{3} \GeV^{-2}$
                                   &  0.82 + 1.65 &  0.23 + 1.46 & -1.92 + 0.62 & -1.47 + 0.83 \\
\hline
$\coeffv{0}{2}$ & $\times 10^{3} \GeV^{-4}$
                                   &  3.15 + 0.39 &  3.21 + 0.49 &  3.07 + 0.42 &  2.89 + 0.30 \\
\hline
$\coeffv{1}{2}$ & $\times 10^{3} \GeV^{-2}$
                                   & -0.16 + 0.34 & -0.12 + 0.32 & -0.07 + 0.17 & -0.13 + 0.13 \\
\hline
$\cbeta{0}{0}$  & $\times 10^{3}$  & -0.12 + 0.18 & -0.07 + 0.22 & -0.06 + 0.19 & -0.02 + 0.32 \\
\hline
$\cbeta{1}{0}$  & $\times 10^{3}$  & -0.47 + 0.25 & -0.57 + 0.22 & -0.49 + 0.14 & -0.49 + 0.18 \\
\hline
$\cbeta{2}{0}$  & $\times 10^{3}$  &  0.77 + 0.18 &  0.72 + 0.23 &  0.69 + 0.19 &  0.64 + 0.34 \\
\hline
$\cbeta{3}{0}$  & $\times 10^{3}$  &  0.05 + 0.55 &  0.19 + 0.51 &  0.06 + 0.26 &  0.11 + 0.15 \\
\hline
$\cbeta{4}{0}$  & $\times 10^{3}$  &  0.33 + 0.18 &  0.34 + 0.15 &  0.39 + 0.12 &  0.42 + 0.14 \\
\hline
$\cbeta{5}{0}$  & $\times 10^{3}$  &  0.73 + 0.62 &  0.86 + 0.66 &  0.77 + 0.83 &  0.48 + 0.22 \\
\hline
$\cbeta{6}{0}$  & $\times 10^{3}$  &  0.00 + 0.30 &  0.06 + 0.28 & -0.25 + 0.13 & -0.34 + 0.15 \\
\hline
$\cbeta{7}{0}$  & $\times 10^{3}$  &  0.13 + 0.25 &  0.42 + 0.33 &  1.01 + 0.49 &  0.76 + 0.48 \\
\hline
$\cbeta{8}{0}$  & $\times 10^{3}$  & -0.06 + 0.41 & -0.38 + 0.46 &  0.02 + 0.51 &  0.15 + 0.14 \\
\hline
$\cbeta{13}{0}$ & $\times 10^{3}$  & -0.08 + 0.65 & -0.02 + 0.60 &  0.16 + 0.42 &  0.23 + 0.17 \\
\hline
$\cbeta{14}{0}$ & $\times 10^{3}$  &  0.08 + 0.35 & -0.03 + 0.31 & -0.21 + 0.22 & -0.25 + 0.14 \\
\hline
$\cbeta{18}{0}$ & $\times 10^{3}$  &  0.80 + 0.80 &  0.99 + 0.82 &  1.51 + 0.47 &  1.49 + 0.30 \\
\hline
$a_{\pi \pi}^{I=J=1}$ & $\times 10^{2}$ 
                                   & -6.1  + 0.3  & -6.1  + 0.2  & -6.1  + 0.3  & -6.0  + 0.2  \\
\hline \hline
\multicolumn{2}{|c|}{$R_{0 \eta}$} &  0.13 + 0.26 &  0.24 + 0.23 &  0.61 + 0.13 &  0.55 + 0.17 \\
\hline
\multicolumn{2}{|c|}{$R_{8 \eta'}$}&  0.22 + 0.11 &  0.20 + 0.08 & -0.01 + 0.11 & -0.04 + 0.08 \\
\hline
\end{tabular}
\caption{Numerical values of the fit parameters itemized according to the four different clusters
         of fits. They also determine the two $\eta$-$\eta'$ mixing parameters $R_{0 \eta}$
         and $R_{8 \eta'}$. The regularization scale in $G$ is set to $\mu = 1$~GeV.}
\label{tab:LECs}
\end{table}

\subsection{Recent experimental developments} \label{sec:newexp}

Very recently the Dalitz plot distributions of the decays $\eta \to 3\pi$ and $\eta' \to \eta \pi^+ \pi^-$
have been determined experimentally with high statistics by the KLOE \cite{KLOE} and the VES Collaboration
\cite{VES}, respectively. In this section we will discuss the changes of our results when these new
and precise (though not yet published) data are included in the fit instead of the PDG values.
In Table~\ref{tab:VES} we show the results of a fit, where the VES numbers are taken into 
account.\footnote{Note, however, that the analysis of the VES experiment is still not completed
and the quoted numbers may slightly change.}
Since the amplitudes for $\eta' \to \eta \pi^+ \pi^-$ and $\eta' \to \eta \pi^0 \pi^0$ would be equal in the
isospin limit and deviations are thus isospin-breaking and small in our approach, we only include the leading
Dalitz parameter $a$ of $\eta' \to \eta \pi^0 \pi^0$ and omit the higher ones which 
are---assuming only small isospin-violating contributions---not quite compatible
with the new results of the VES experiment for $\eta' \to \eta \pi^+ \pi^-$.  Within our approach 
the $c$ value has a tendency to remain on the positive side in contrast to the result of the VES Collaboration,
nevertheless our results are in reasonable overall agreement with the Dalitz plot parameters extracted 
from the VES experiment. Most remarkably, none of the various other observables (decay widths, branching 
ratios, Dalitz parameters of other decay modes, etc.) is significantly altered when the VES numbers are
included in the fit, so that the very good agreement of the results with all published data of hadronic 
$\eta$ and $\eta'$ decays is retained.

\begin{table}
\centering
\begin{tabular}{|l|D{+}{\,\pm\,}{6,6}|D{+}{\,\pm\,}{6,6}|D{+}{\,\pm\,}{6,6}|}
\hline
\multicolumn{4}{|c|}{$\eta' \to \eta \pi^+ \pi^-$} \\
\hline
& \multicolumn{1}{c|}{$a$} &
\multicolumn{1}{c|}{$b$} &
\multicolumn{1}{c|}{$c$} \\
\hline
theo. &
 -0.116  +   0.011  &
 -0.042  +   0.034  &
  0.010  +   0.019  \\
\hline
exp.\ \cite{VES}&
 -0.132  +   0.019  &
 -0.108  +   0.033  &
 -0.046  +   0.022  \\
\hline\hline
\multicolumn{4}{|c|}{$\eta' \to \eta \pi^0 \pi^0$} \\
\hline
& \multicolumn{1}{c|}{$a$} &
\multicolumn{1}{c|}{$b$} &
\multicolumn{1}{c|}{$c$} \\
\hline
theo. &
 -0.127  +   0.009  &
 -0.049  +   0.036  &
  0.011  +   0.021  \\
\hline
exp.\ \cite{pdg} &
 -0.116  +   0.026  &
\multicolumn{1}{c|}{ } &
\multicolumn{1}{c|}{ } \\
\hline
\end{tabular}
\caption{Results for the Dalitz plot parameters of $\eta' \to \eta \pi \pi$ when the VES data \cite{VES}
         are included in the fit.}
\label{tab:VES}
\end{table}

Next, we have replaced the Dalitz plot parameters of $\eta \to \pi^+ \pi^- \pi^0$ and $\eta \to 3 \pi^0$
quoted in Sec.~\ref{sec:83xx} by the new and precise results of the KLOE Collaboration \cite{KLOE}
omitting again the VES numbers, in order to avoid interference of these two new, but so far unpublished data 
sets. 
The results are compiled in Table~\ref{tab:KLOE}. While it is possible to accommodate the KLOE numbers for
the $a$ and $c$ coefficients of the $\eta \to \pi^+ \pi^- \pi^0$ Dalitz plot distribution, our results do 
not agree with $b$ and $d$. In particular, the value of the $y^2$-coefficient $b$ differs from the KLOE number,
which has been determined very precisely, by more than five standard deviations. Within the given 
boundaries for the low-energy coefficients of the chiral effective Lagrangian our approach is unable to
produce a $b$ value as small as the number advocated by the KLOE Collaboration \cite{KLOE}. Note that such
a small value also implies unexpectedly large corrections to the well-known current algebra result $b = a^2/4$
\cite{GL2, OW}.
It may indicate that contributions from higher chiral orders of the effective Lagrangian
could play a role for this quantity. But the inclusion of such higher orders
is beyond the scope of the present investigation and will not be discussed here.

On the other hand, the KLOE result for the leading order coefficient of the $\eta \to 3 \pi^0$ Dalitz plot,
$g$, cannot be met, while our result still remains compatible with the PDG value, $-0.062 \pm 0.008$ 
\cite{pdg}. Generally, we observe the pattern, that reducing the $b$ value correlates with an enhancement
of the modulus of $g$. Finally, fitting to the new KLOE numbers destroys the agreement of the measured 
branching ratio $r_1$, Eq.~(\ref{eq:defBR}), and our result, which is significantly increased. 
The accordance of the rest of the 
calculated observables with experimental data is only marginally affected by including the KLOE results, 
also the partial decay widths of the two $\eta \to 3 \pi$ decay modes which enter $r_1$ remain consistent
with the---admittedly large---experimental error bars.

In Table~\ref{tab:KLOE} we employ the $r_1$ value which is determined by the Particle Data Group 
by performing a $\chi^2$-fit using one decay rate and 18 branching ratios 
(quoted as ``our fit'' in \cite{pdg}). The result of the most recent 
direct measurement of $r_1$ \cite{CMD}, however, is a bit larger and has also larger error bars: 
$r_1 = 1.52 \pm 0.12$, where we have added statistical and systematic errors linearly. Employing this number 
instead of the PDG value slightly improves the fit to the KLOE data, but does not resolve the disagreement
with the Dalitz parameters $b$ and $g$. Taking this value for $r_1$
we find $a = -1.049 \pm 0.025$, $b = 0.178 \pm 0.019$, 
$c = 0.079 \pm 0.028$, $d = 0.064 \pm 0.012$, $g = -0.056 \pm 0.012$, $r_1 = 1.51 \pm 0.01$.

We mention in passing that after relaxing the naturalness assumption on the size of the chiral parameters
described at the beginning of this section, we have found a second class of fits, which are slightly 
closer to the results of the KLOE Collaboration for the Dalitz plot of $\eta \to \pi^+ \pi^- \pi^0$.
Apart from involving unnaturally large values of some of the LECs they entail a $g$ value which
is even larger in magnitude than the one of the previous fits, Table~\ref{tab:KLOE}. Moreover, the 
agreement with the experimental phase shifts of $\pi \pi$ scattering in the $I=J=0$ channel
shown in Fig.~\ref{fig:PhShPlots} is 
considerably worsened. The branching ratio $r_1$, on the other hand, is not altered, cf.\ Table~\ref{tab:KLOE}.

\begin{table}
\centering
\begin{tabular}{|l|D{+}{\,\pm\,}{6,6}|D{+}{\,\pm\,}{6,6}|D{+}{\,\pm\,}{5,5}|D{+}{\,\pm\,}{5,5}|}
\hline
\multicolumn{5}{|c|}{$\eta \to \pi^+ \pi^- \pi^0$} \\
\hline
& \multicolumn{1}{c|}{$a$ \cite{KLOE}} &
\multicolumn{1}{c|}{$b$ \cite{KLOE}} &
\multicolumn{1}{c|}{$c$ \cite{KLOE}} &
\multicolumn{1}{c|}{$d$ \cite{KLOE}} \\
\hline
theo. &
 -1.054  +   0.025  &
  0.185  +   0.015  &
  0.079  +   0.026  &
  0.064  +   0.012  \\
\hline
exp.  &
 -1.072  +   0.013  &
  0.117  +   0.012  &
  0.047  +   0.011  &
  0.13   +   0.03   \\
\hline
\hhline{===}
& \multicolumn{2}{c|}{$\eta \to 3 \pi$} \\
\hhline{---}
& \multicolumn{1}{c|}{$g$ \cite{KLOE}} &
\multicolumn{1}{c|}{$r_1$ \cite{pdg}} \\
\hhline{---}
theo. &
 -0.058  +   0.011  &
  1.50   +   0.01   \\
\hhline{---}
exp.  &
 -0.026  +   0.018  &
  1.44   +   0.04   \\
\hhline{---}
\end{tabular}
\caption{Results for the Dalitz plot parameters of $\eta \to  3\pi$ and the branching ratio $r_1$, 
         Eq.~(\ref{eq:defBR}), when the KLOE data \cite{KLOE} are included in the fit.
         For simplicity we have added the statistical and systematic errors specified in 
         \cite{KLOE} linearly and display symmetrized error bars according to the larger value.
         Note that our coefficients $c$, $d$, and $g$ correspond to $d$, $f$, and $2\alpha$ in 
         \cite{KLOE}, respectively.}
\footnotetext{For simplicity we have added the statistical and systematic errors specified in 
              \cite{KLOE} linearly and display symmetrized error bars according to the larger value.}%
\label{tab:KLOE}
\end{table}

\section{Dalitz plot parameters of \mbox{\boldmath $\eta \to 3 \pi$}} \label{sec:DalPhe}

As pointed out in the previous subsection it is not possible to accommodate the new KLOE results for the 
Dalitz parameters of $\eta \to 3\pi$ together with the measured branching ratio of the two
decay modes, $r_1$. In this subsection we will present an explanation how all these experimental quantities 
can be related in a phenomenological way without making use of model-dependent assumptions on the 
construction of the decay amplitudes.

The main ingredient is the $\Delta I = 1$ selection rule which relates the $\eta \to \pi^+ \pi^- \pi^0$ 
decay amplitude $\mathcal{A}$ to the amplitude $\bar{\mathcal{A}}$ for $\eta \to 3 \pi^0$ \cite{GL2}
\beq \label{eq:DeltaI}
\bar{\mathcal{A}}(s,t,u) = \mathcal{A}(s,t,u) + \mathcal{A}(t,u,s) + \mathcal{A}(u,s,t) \,.
\eeq
This rule is valid up to tiny corrections from QCD (suppressed by $\mathcal{O}(m_{\epsilon}^2)$) 
and of electromagnetic origin (suppressed by $\mathcal{O}(\alpha^2)$).\footnote{Our chiral unitary approach 
iterates isospin-breaking terms and thus includes corrections to the $\Delta I = 1$ selection rule, 
but we have checked that these are numerically tiny.}
In analogy to the experimental parametrization of the Dalitz plot distribution, Eq.~(\ref{eq:DalC}), we 
assume that the amplitude $\mathcal{A}$ can be well approximated by a polynomial
\beq \label{eq:Apoly}
\mathcal{A}(x,y) = N \big[1 + \alpha y + \beta y^2 + \gamma x^2 + \cdots \big]
\eeq
with complex coefficients $\alpha$, $\beta$, $\gamma$. We will drop all terms of third order and beyond 
and work with this minimal parametrization of $\mathcal{A}$ which is able to 
describe the experimental Dalitz plot distribution as
measured by the KLOE Collaboration\footnote{Note that in contrast to \cite{KLOE} we do not 
include $C$-violating terms proportional to $x$. Therefore, our coefficients $c$ and $d$ correspond to $d$ 
and $f$ in \cite{KLOE}, respectively.}
\cite{KLOE}
\beq \label{eq:DalKLOE}
\begin{array}{c}
|\mathcal{A}(x,y)|^2  = |N|^2 \big[ 1 + a y + b y^2 + c x^2 + d y^3 ] \\[1.0ex]
\textrm{with}\footnotemark
  \quad a = -1.072 \pm 0.013, \quad b = 0.117 \pm 0.012, 
  \quad c =  0.047 \pm 0.011, \quad d = 0.13  \pm 0.03 \,.
\end{array}
\eeq
\footnotetext{For simplicity we have added the statistical and systematic errors specified in 
              \cite{KLOE} linearly and display symmetrized error bars according to the larger value.}%
Employing the $\Delta I = 1$ selection rule, Eq.~(\ref{eq:DeltaI}), we are able to derive expressions 
for the leading Dalitz plot parameter of $\eta \to 3\pi^0$, $g$, and for the branching ratio $r_1$.
Since the complex normalization factor $N$ is irrelevant for
the determination of the $g$ parameter and drops out in the branching ratio $r_1$, 
we are left with six free constants which parametrize the amplitude $\mathcal{A}$, the real and 
imaginary parts of $\alpha$, $\beta$, and $\gamma$. Four of these can be fixed by matching $|\mathcal{A}|^2$
to the central experimental values of $a$, $b$, $c$, $d$. However, also the remaining two are constrained 
by the fact that the higher order terms $x^2 y$, $x^2 y^2$, $x^4$, and $y^4$, which automatically emerge 
when squaring $\mathcal{A}$, are expected to have small coefficients,
since Eq.~(\ref{eq:DalKLOE}) appears to be a good parametrization of the experimental distribution.
As an upper limit for the moduli of these higher coefficients not observed in experiment we choose 
the value of the highest order experimental coefficient in Eq.~(\ref{eq:DalKLOE}), $d = 0.13$.

Fitting the remaining two parameters in $\mathcal{A}$ within the boundaries dictated by the smallness of 
the higher order terms in $|\mathcal{A}|^2$ to the experimental numbers for $r_1$ and KLOE $g$ we find
\beq \label{eq:phefit}
\begin{array}{rcrclrcrcll}
g^{\textrm{(theo)}}      &=& -0.074 &\pm& 0.012,  \mbox{\qquad} 
& g^{\textrm{(exp)}}     &=& -0.026 &\pm& 0.018 & \mbox{\cite{KLOE}}, \\[1.0ex]
r_{1}^{\textrm{(theo)}}  &=&  1.47  &\pm& 0.03,   \mbox{\qquad} 
& r_{1}^{\textrm{(exp)}} &=&  1.44  &\pm& 0.04  & \mbox{\cite{pdg}},
\end{array}
\eeq
where the theoretical uncertainties represent the propagation of the errors of the input parameters in 
Eq.~(\ref{eq:DalKLOE}). While the two numbers for the branching ratio $r_1$ are very well compatible,
the calculated value of $g$ differs by about two standard deviations from the number extracted by the KLOE 
Collaboration. It is, however, consistent with the value published by the Crystal Ball Collaboration 
\cite{Tippens}.

We would like to point out that raising the order of the polynomial parametrization of the amplitude
in Eq.~(\ref{eq:Apoly}) does not alter these conclusions. Although it would increase the number of adjustable
parameters, at the same time more and more constraints would be generated by the fact that the numerous 
higher order coefficients of $|\mathcal{A}|^2$ all have to be close to zero for Eq.~(\ref{eq:DalKLOE}) to 
be a good parametrization of the experimental distribution. As a matter of fact, 
we have explicitly checked that the inclusion of, e.g.,
a $y^3$ term in the parametrization of the amplitude, Eq.~(\ref{eq:Apoly}), yields only tiny numerical 
improvements for the fit to $g$ and $r_1$.
As in Sec.~\ref{sec:newexp}, we have verified that our results do not change significantly, when the 
PDG value for $r_1$ is replaced by the most recent direct experimental determination of this branching ratio
which yields $r_1 = 1.52 \pm 0.12$ \cite{CMD}. Instead of the numbers given in Eq.~(\ref{eq:phefit}) 
we then obtain $g = -0.071 \pm 0.012$ which is only slightly closer to the KLOE number and 
$r_1 = 1.50 \pm 0.03$. The main restriction for the parameters is thus given by the size of the higher 
order coefficients of $|\mathcal{A}|^2$ and not by the value of $r_1$.

We have also checked to what extent the phenomenological amplitude described here fulfills the unitarity 
condition discussed in Sec.~\ref{sec:uni}. This can be done utilizing purely experimental input and thus 
without making use of unitarized ChPT, since the scattering amplitude $\hat{T}_l$ which enters 
Eq.~(\ref{eq:unired}) may be expressed by the experimentally determined phase shifts of $\pi \pi$ 
scattering, see \cite{Walker} for the explicit expressions. The violation of three-particle unitarity
turns out to be 10\,\% (5\,\%) for $\eta \to \pi^+ \pi^- \pi^0$ ($\eta \to 3 \pi^0$) on average over the 
full Dalitz plot when the parameters are fixed to physical observables as above, cf.\ 
Eqs.~(\ref{eq:DalKLOE}, \ref{eq:phefit}). 
The free normalization constant $N$ is chosen in such a way that the unitarity violation is minimized 
at the center of the $\eta \to \pi^+ \pi^- \pi^0$ Dalitz plot.
Although the polynomial amplitude does not incorporate any 
constraints from unitarity the violations turn out to be rather modest.
If, on the other hand, the restrictions on the size of higher order coefficients of $|\mathcal{A}|^2$ are 
released and the fit is forced to reproduce the central values of the branching ratio $r_1$ and the KLOE $g$ 
value, we observe unitarity violations as large as 43\,\% for $\eta \to \pi^+ \pi^- \pi^0$ and 45\,\% for 
$\eta \to 3 \pi^0$.

\section{Conclusions}  \label{sec:con}

In the present work we have investigated the hadronic decays 
$\eta, \eta' \to 3 \pi$ and $\eta' \to \eta \pi \pi$ within a chiral unitary
approach based on the Bethe-Salpeter equation. The $s$- and $p$-wave
interaction kernels of the BSE are derived
from the U(3) chiral effective Lagrangian up to fourth chiral order with
the $\eta'$ as an explicit degree of freedom.
Within this approach the incoming $\eta$ or $\eta'$ decays into three pseudoscalar mesons
and then two of these mesons rescatter---elastically or inelastically---an arbitrary
number of times, while the third meson remains a spectator. The final state
interaction of the two mesons is described by the solution of the BSE
and satisfies two-particle unitarity. For the decays $\eta \to 3\pi$ and
$\eta' \to \eta \pi \pi$ we have also estimated to
what extend constraints from three-body unitarity, which is not incorporated in the 
approach, are fulfilled and find that the deviations are rather modest.

The chiral parameters of the approach are fitted by means of an overall
$\chi^2$ fit to available data on
the hadronic decay modes of $\eta$ and $\eta'$ and meson-meson scattering
phase shifts. We obtain very good agreement with currently available data
on the decay widths and spectral shapes. In fact, we observe four different classes
of fits which describe these data equally well, but differ in their predictions for yet unmeasured
quantities such as the $\eta' \to \pi^+ \pi^- \pi^0$ decay width (for which there exists
only a weak upper limit) and the Dalitz slope parameters of $\eta' \to 3 \pi$.
The results obtained may be tested in future experiments foreseen at WASA@COSY and MAMI-C.
The hadronic decays considered here along with phase shifts in meson-meson
scattering pose therefore tight constraints on the approach and will allow to 
determine the couplings of the effective Lagrangian up to fourth chiral order.
It is important to stress that the values of the parameters obtained from the fit
are in general not the same as in the framework of ChPT which can be traced back to the
absorption of loops into the coefficients and higher order effects not included
in the perturbative framework.

An intriguing feature of the fits is that they accommodate the large negative
slope parameter $g$ of the decay $\eta \to 3 \pi^0$ measured by the 
Crystal Ball Collaboration \cite{Tippens} which could not be met by previous
theoretical investigations. This value must, however, be confronted with the more recent
but yet preliminary $g$ value of the KLOE Collaboration \cite{KLOE}.
If we replace the PDG data by the KLOE Dalitz parameters of both the charged and neutral
$\eta \to 3 \pi$ decay, we do not achieve
a good overall fit. It appears that the slope parameters of both $\eta \to 3 \pi$ decays
cannot be fitted simultaneously. In addition, fitting to the KLOE data
destroys the agreement with the experimental branching ratio
of both decays which is known to high precision. To this end, we have illustrated that
utilizing the $\Delta I =1$ selection rule which relates both decays and taking 
the KLOE parametrization of the charged decay as input
leads in a model-independent way to a $g$ value not consistent with the KLOE $g$ result.

The importance of the various two-particle channels with different isospin
and angular momentum has been examined as well. For the $\eta \to 3 \pi$ decays we find that the major
contribution is given by $\pi \pi$ rescattering in the $s$-wave $I=0$ channel,
while the $I=1,2$ channels interfere destructively with the former. The $p$-wave
contribution in the charged decay is tiny, since available phase space is small 
and the $C$-odd channels related to the $\rho(770)$ resonance do not occur.
For $\eta' \to \pi^+ \pi^- \pi^0$, on the other hand, phase space is considerably
larger, and the size of the $p$-wave contributions ranges from 10\,\% to 50\,\%
depending on the cluster of fits.

For the decays $\eta' \to \eta \pi \pi$ we find that the
$s$-wave $I=1$ channels dominate for two classes of fits which
would confirm the importance of the nearby $a_0(980)$ resonance as claimed
by previous investigations. But the other two clusters are dominated
by the $I=0$ channels. These two scenarios can be distinguished by their predictions
for the $\eta' \to 3 \pi$ decays. Thus, a precise measurement of $\eta' \to 3 \pi$ decay 
parameters can also help to clarify the importance of $a_0(980)$ or $f_0(980)$ resonance contributions 
to the dominant decay mode of the $\eta'$ into $\eta \pi \pi$.

\section*{Acknowledgements}

We thank F.~Ambrosino, B.~Di~Micco, J.~Gasser, B.~Nefkens, V.~Nikolaenko, A.~Starostin 
and M.~Wolke for useful discussions.
This work was supported in part by DFG, SFB/TR-16 ``Subnuclear Structure of Matter'', and
Forschungszentrum J\"ulich.

\appendix

\section{Fourth-order operators} \label{sec:AppLagr}

For completeness, we tabulate those pieces of the Lagrangian of fourth chiral order which are employed
in this work. The full list can be found in \cite{H-S, BB3}. The fourth-order Lagrangian is of the form
\beq
\Lagr^{(4)} = \sum_i \beta_i \ \mathcal{O}_i \,,
\eeq
where the $\beta_i$ are functions of the singlet field $\eta_0$ which can be expanded in terms of this 
variable in the same manner as the $V_i$ in Eq.~(\ref{eq:Lagr02}) with expansion coefficients 
$\cbeta{i}{j}$.  The operators $\mathcal{O}_i$ which are relevant for this work read
\beq
\begin{array}{lcllcl}
\mathcal{O}_{0}  & = & \trf{C^{\mu} C^{\nu} C_{\mu} C_{\nu}}, & \qquad
\mathcal{O}_{1}  & = & \trf{C^{\mu} C_{\mu}} \trf{C^{\nu} C_{\nu}},  \\
\mathcal{O}_{2}  & = & \trf{C^{\mu} C^{\nu}} \trf{C_{\mu} C_{\nu}}, & \qquad
\mathcal{O}_{3}  & = & \trf{C^{\mu} C_{\mu} C^{\nu} C_{\nu}},  \\
\mathcal{O}_{4}  & = & -\trf{C^{\mu} C_{\mu}} \trf{M}, & \qquad
\mathcal{O}_{5}  & = & -\trf{C^{\mu} C_{\mu} M}, \\
\mathcal{O}_{6}  & = & \trf{M} \trf{M}, & \qquad
\mathcal{O}_{7}  & = & \trf{N} \trf{N}, \\
\mathcal{O}_{8}  & = & \frac{1}{2} \trf{M M + N N}, & \qquad
\mathcal{O}_{13} & = & -\trf{C^{\mu}} \trf{C_{\mu} C^{\nu} C_{\nu}}, \\
\mathcal{O}_{14} & = & -\trf{C^{\mu}} \trf{C_{\mu}} \trf{C^{\nu} C_{\nu}}, & \qquad 
\mathcal{O}_{18} & = & -\trf{C^{\mu}} \trf{C_{\mu} M},
\end{array}
\eeq
where we have made use of the abbreviations
\beq
C_{\mu} = U^{\dagger} \partial_{\mu} U, \qquad
M = U^{\dagger} \chi + \chi^{\dagger} U, \qquad
N = U^{\dagger} \chi - \chi^{\dagger} U.
\eeq

\section{Phase shifts of meson-meson scattering} \label{sec:AppPhSh}

\begin{figure}
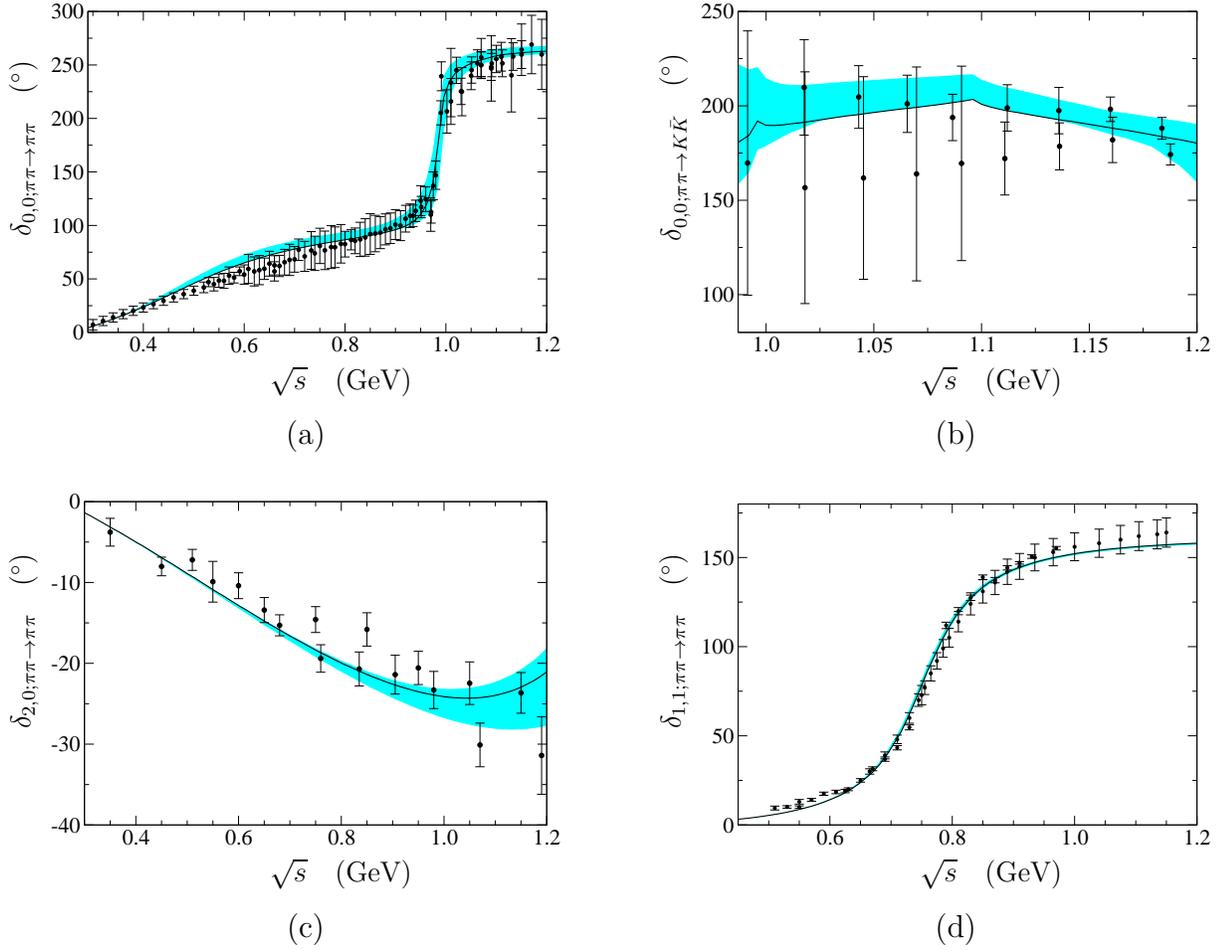

\centering
\begin{tabular}{cp{1.0cm}c}
\begin{overpic}[width=0.4\textwidth,clip]{PhSh00pp.eps}
  \put(43,-7){\scalebox{0.9}{$\sqrt{s}$\quad (GeV)}}
  \put(-8,23){\rotatebox{90}{\scalebox{0.9}{$\delta_{0,0;\pi\pi \to \pi\pi}\quad (^\circ)$}}}
\end{overpic} & &
\begin{overpic}[width=0.4\textwidth,clip]{PhSh00pK.eps}
  \put(43,-7){\scalebox{0.9}{$\sqrt{s}$\quad (GeV)}}
  \put(-8,22){\rotatebox{90}{\scalebox{0.9}{$\delta_{0,0;\pi\pi \to K \bar{K}}\quad (^\circ)$}}}
\end{overpic} \\[0.7cm]
(a) & & (b) \\[0.5cm]
\begin{overpic}[width=0.4\textwidth,clip]{PhSh20pp.eps}
  \put(43,-7){\scalebox{0.9}{$\sqrt{s}$\quad (GeV)}}
  \put(-8,23){\rotatebox{90}{\scalebox{0.9}{$\delta_{2,0;\pi\pi \to \pi\pi}\quad (^\circ)$}}}
\end{overpic} & &
\begin{overpic}[width=0.4\textwidth,clip]{PhSh11pp.eps}
  \put(43,-7){\scalebox{0.9}{$\sqrt{s}$\quad (GeV)}}
  \put(-8,23){\rotatebox{90}{\scalebox{0.9}{$\delta_{1,1;\pi\pi \to \pi\pi}\quad (^\circ)$}}}
\end{overpic} \\[0.7cm]
(c) & & (d) \\
\end{tabular}
\caption{Results for the phase shifts $\delta_{I,J}$ of meson-meson scattering for isospin $I$
         and partial wave $J$. The shaded area indicates the range of fits taken
         into account within our approach, while the solid line represents the best fit in each 
         particular channel. 
         The data are from \cite{I0J0pipi} (a), \cite{I0J0piK} (b), 
         \cite{I2J0pipi} (c), and \cite{I1J1pipi} (d).}
\label{fig:PhShPlots}
\end{figure}

In Fig.~\ref{fig:PhShPlots} we show the results for the phase shifts of four meson-meson channels:
the $I=0,2$ $s$-wave and $I=1$ $p$-wave $\pi \pi \to \pi \pi$ scattering as well as 
$\pi \pi \to K \bar{K}$ with $I=J=0$. 
The agreement with the experimental data points is remarkably good.
The shaded areas indicate the variation of the results when the overall $\chi^2$ value of the fits is 
allowed to exceed the minimum by at most 15\,\%, cf.\ Sec.~\ref{sec:res}.
The variation is particularly small for $I=J=1$ $\pi \pi$ scattering which entails the $\rho$(770) 
resonance. This fact is, however, not surprising since this channel does not enter the hadronic $\eta$
and $\eta'$ decays and involves an additional fit parameter, the subtraction constant 
$a_{\pi\pi}^{(I=J=1)}$.

\end{document}